\documentclass[acmsmall]{acmart}
\usepackage{enumitem}
\usepackage{amsmath}
\usepackage{algorithm}
\usepackage{algorithmic}
\usepackage{xspace}
\usepackage{xcolor,colortbl}
\usepackage{booktabs}
\usepackage{tabularx} %
\usepackage{float}
\usepackage{subcaption}
\usepackage{graphicx}
\usepackage{makecell}
\usepackage{multirow}
\usepackage{placeins}
\usepackage{makecell}
\usepackage{pifont}
\usepackage{wrapfig}

\newcommand{\cmark}{\ding{51}} %
\newcommand{\xmark}{\ding{55}} %
\newcommand{\omark}{\ding{109}} %

\begin{document}

\setcopyright{cc}
\setcctype{by}
\acmJournal{POMACS}
\acmYear{2026} \acmVolume{10} \acmNumber{3} \acmArticle{67}
\acmMonth{12} \acmDOI{10.1145/3857370}

\received{10 July 2026}
\received[revised]{5 October 2026}
\received[accepted]{9 October 2026}

\title{
Is Your NPU Ready for LLMs? Dissecting the Hidden Efficiency Bottlenecks in Mobile LLM Inference}

\author{Guanyu Cai}
\email{caigy25@mails.tsinghua.edu.cn}
\affiliation{%
  \institution{Tsinghua University}
  \city{Beijing}
  \country{China}
}

\author{Ruiming Tian}
\email{22301131@bjtu.edu.cn}
\affiliation{%
  \institution{Beijing Jiaotong University}
  \city{Beijing}
  \country{China}
}

\author{Lang Yang}
\email{yanglang21@mails.tsinghua.edu.cn}
\affiliation{%
  \institution{Tsinghua University}
  \city{Beijing}
  \country{China}
}

\author{Zhouhong Ren}
\email{23126462@bjtu.edu.cn}
\affiliation{%
  \institution{Beijing Jiaotong University}
  \city{Beijing}
  \country{China}
}

\author{Jinliang Yuan}
\email{yuanjinliang@tsinghua.edu.cn}
\affiliation{%
  \institution{Tsinghua University}
  \city{Beijing}
  \country{China}
}

\author{Lingkun Li}
\email{lkli@bjtu.edu.cn}
\affiliation{%
  \institution{Beijing Jiaotong University}
  \city{Beijing}
  \country{China}
}

\author{Jiliang Wang}
\authornote{Corresponding author: Jiliang Wang (jiliangwang@tsinghua.edu.cn).}
\email{jiliangwang@tsinghua.edu.cn}
\affiliation{%
  \institution{Tsinghua University}
  \city{Beijing}
  \country{China}
}

\begin{abstract}

Deploying Large Language Models (LLMs) on mobile devices enhances privacy and reduces latency, but is severely constrained by hardware inefficiencies.
We present the first comprehensive, cross-layer measurement study of mobile LLM inference, spanning five mainstream frameworks (e.g., llama.cpp and GENIE) and three hardware backends (CPU, GPU, and NPU).
To enable this analysis, we develop PowerBench, a fine-grained profiling tool that measures LLM throughput and provides the first backend-specific energy attribution for mobile LLM inference, going beyond traditional device-level measurements. We further conduct hardware microbenchmarks to characterize the memory-system performance and compute capabilities of mobile CPUs, GPUs, and NPUs.
Our study yields three critical insights: (1) Framework-induced performance gaps are substantially amplified on NPUs, reaching up to $10\times$ due to differences in offloading strategies, hardware-underutilizing operators, and activation quantization strategies.
(2) We identify a distinct ``phase split,'' in which NPUs excel at compute-bound prefilling, while CPUs outperform all other backends in memory-bound decoding. This split is driven by NPUs' preference for large, fixed-shape workloads, which conflicts with the small-kernel, dynamic nature of decoding.
(3) Backend-specific profiling uncovers substantial scheduling headroom missed by prior work. Suboptimal thread configurations, uncoordinated NPU sleep latencies, and CPU polling intervals result in unnecessary energy consumption. For example, tuning CPU frequency alone reduces energy consumption by up to 39.7\%.
Leveraging these findings, we estimate that an energy-oriented configuration can reduce energy consumption by up to 53.2\% on the NPU backend across three datasets, highlighting the substantial potential for further efficiency improvements in mobile LLM deployment.

\end{abstract}
\keywords{Mobile LLM, Performance Measurement, Energy Efficiency}

\ccsdesc[500]{General and reference~Measurement}
\ccsdesc[500]{General and reference~Performance}
\ccsdesc[500]{Human-centered computing~Mobile devices}
\ccsdesc[500]{Computing methodologies~Natural language generation}
\ccsdesc[500]{Human-centered computing~Mobile computing}

\maketitle

\section{Introduction}

Large Language Models (LLMs) have demonstrated remarkable capabilities in language understanding, reasoning, and multimodal interaction~\cite{chung_scaling_2024,liu_visual_2023}.
Deploying LLMs on mobile devices is becoming increasingly attractive, motivated by the need for lower latency, stronger privacy, and offline accessibility~\cite{xu_-device_2024_review}.
This is particularly important for LLM-powered mobile agents, which sustain interactive inference while invoking tools and manipulating user interfaces~\cite{wenAutoDroidV2BoostingSLMbased2025,wen_autodroid_2024}.
Sustaining agent workflows demands both high throughput for interactive responsiveness and low energy consumption to preserve battery life, making efficiency the central challenge.
This challenge has motivated a growing body of optimization efforts across the LLM inference stack, including inference frameworks~\cite{xuFastOndeviceLLM2024, xue2024powerinfer,EdgeMoE}, execution backends~\cite{xuFastOndeviceLLM2024,chen_characterizing_2025}, and resource scheduling~\cite{huangMNNAECSEnergyOptimization2025,zhangDissectingImpactMobile2025a}.

However, the current research landscape remains fragmented, as existing works typically focus on isolated components of the LLM inference stack. This fragmentation limits the community's ability to develop a holistic understanding of mobile LLM efficiency. While recent measurement studies \cite{zhangDissectingImpactMobile2025a,luDemystifyingSmallLanguage2025,li_large_2024,guo_large_2025,benazir2025benchmarkingapplesilicon} have demonstrated the feasibility of on-device LLMs---characterizing the impact of model scale, quantization, and CPU/GPU backends---they leave two critical blind spots. 
\begin{wrapfigure}[11]{r}{0.45\linewidth}
    \centering
    \vspace{-6pt}
    \includegraphics[width=\linewidth]{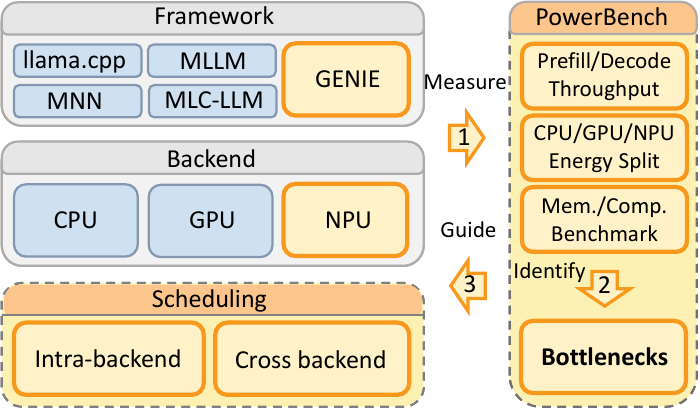}
    \captionsetup{skip=1pt}
    \caption{Cross-layer measurement overview.}
    \label{fig:introduction}
    \vspace{-4pt}
\end{wrapfigure}
First, despite NPUs emerging as a primary target for deployment, their performance characteristics across diverse frameworks remain poorly understood. 
Second, the synergistic impact of resource scheduling across heterogeneous backends on latency and energy consumption remains largely underexplored. Table~\ref{tab:comparison} compares the research coverage of our work with that of prior studies.

To address these blind spots, we conduct a comprehensive cross-layer measurement study of mobile LLM inference, as shown in Figure~\ref{fig:introduction}. It is the first to simultaneously investigate NPU execution and resource scheduling. Our study encompasses five mainstream frameworks (e.g., llama.cpp and GENIE), three heterogeneous backends (CPU, GPU, and NPU), and diverse resource scheduling policies (e.g., DVFS).

\textbf{Measurement methodology:} Our analysis is underpinned by a unified measurement framework. To quantify energy efficiency, we develop \texttt{PowerBench}, a lightweight, framework-agnostic instrumentation library. Unlike prior coarse-grained, device-level profiling, \texttt{PowerBench} enables fine-grained, backend-specific energy attribution by capturing both full-SoC and per-compute-unit consumption.
For throughput measurement, we instrument each framework to standardize workload execution, including precise control over prompt token injection and end-of-sequence token replacement. This ensures that prefill and decode throughput are evaluated under consistent prompt and generation lengths. Collectively, these methods allow us to construct what is, to our knowledge, the first comprehensively controlled benchmark for mobile LLM inference efficiency, featuring over 400 configurations across various models, devices, frameworks, and inference stages. 
For a deeper analysis, we further conduct hardware microbenchmarks to characterize the compute capabilities of mobile CPUs, GPUs, and NPUs, along with the latency, capacity, and bandwidth of their cache and memory hierarchies.

Our results reveal three key findings.

\begin{table*}[t]
\centering
\scriptsize
\setlength{\tabcolsep}{1.5pt}
\renewcommand{\arraystretch}{1.25}
\caption{Comparison of research coverage, scheduling factors, and energy measurement granularity.}
\label{tab:comparison}
\begin{tabular*}{\textwidth}{@{\extracolsep{\fill}}lcccccccccccc@{}}
\toprule
\textbf{Work}
& \multicolumn{5}{c}{\textbf{Framework Coverage}}
& \multicolumn{3}{c}{\textbf{Backend Coverage}}
& \multicolumn{2}{c}{\textbf{Scheduling}}
& \multicolumn{2}{c}{\textbf{Energy Measurement}} \\
\cmidrule(lr){2-6} \cmidrule(lr){7-9} \cmidrule(lr){10-11} \cmidrule(lr){12-13}
& \textbf{llama.cpp}
& \makecell{\textbf{MLC-}\\\textbf{LLM}}
& \textbf{MLLM}
& \textbf{MNN}
& \textbf{GENIE}
& \textbf{CPU}
& \textbf{GPU}
& \textbf{NPU}
& \makecell{\textbf{DVFS}\\\textbf{Impact}}
& \makecell{\textbf{Dispatch}\\\textbf{Impact}}
& \makecell{\textbf{Compute}\\\textbf{Unit}}
& \makecell{\textbf{Whole}\\\textbf{Device}} \\
\midrule

\citet{benazir2025benchmarkingapplesilicon}
& \cmark & -- & -- & -- & --    
& -- & \cmark & --
& -- & --
& -- & -- \\

\citet{luDemystifyingSmallLanguage2025}
& \cmark & -- & -- & -- & --
& \cmark & -- & --
& -- & --
& -- & -- \\

\citet{guo_large_2025}
& \cmark & -- & -- & \cmark & --
& \cmark & \cmark & --
& -- & --
& -- & -- \\

\citet{li_large_2024}
& \cmark & \cmark & \cmark & -- & --
& \cmark & \cmark & --
& -- & --
& -- & -- \\

MELTing Point \cite{laskaridisMELTingPointMobile2024}
& \cmark & \cmark & -- & -- & --
& \cmark & \cmark & --
& -- & --
& -- & \cmark \\

Ours
& \cmark & \cmark & \cmark & \cmark & \cmark
& \cmark & \cmark & \cmark
& \cmark & \cmark
& \cmark & \cmark \\

\bottomrule
\end{tabular*}
\end{table*}

\noindent\textbf{(1) Framework-induced performance gaps are substantially amplified on NPUs.}
While performance diversity across frameworks is expected, we observe that NPU execution significantly exacerbates these disparities. Even frameworks sharing the same vendor-provided backend (e.g., Qualcomm QNN) exhibit clear performance gaps: GENIE achieves 1219.2 tokens/s, outperforming MNN's 700.9 tokens/s. This difference primarily stems from suboptimal offloading strategies, where inter-layer communication and layout conversion introduce significant overhead. The gap widens further because of hardware-underutilizing operator implementations: during prefilling, GENIE reaches 1463.7 tokens/s, surpassing llama.cpp (115.1 tokens/s) by over $10\times$, despite their nearly identical decoding throughput ($\sim$23 tokens/s). For the large matrix multiplications exposed by prefilling, GENIE's QNN MatMul uses the NPU's matrix accelerator, whereas llama.cpp's operator uses only the vector accelerator. Consequently, QNN reduces MatMul compute cycles by over $20\times$ relative to llama.cpp at large input sizes. Furthermore, although activation quantization improves throughput by over $2\times$ and reduces energy consumption by more than 50\%, matched GENIE Llama experiments show PPL increases of 4.7\% to 20.7\% under the unified evaluation protocol.

\noindent\textbf{(2) Backend efficiency exhibits a distinct ``phase split'' between prefilling and decoding.}
The optimal execution backend depends fundamentally on the inference phase. NPU backends dominate the compute-bound prefilling stage, achieving peak throughput exceeding 1,400 tokens/s. However, this ranking reverses during the decoding stage: CPUs typically emerge as the fastest backend (achieving 70 tokens/s), with GPUs following closely, while NPUs consistently lag behind. We attribute this reversal to a fundamental architectural mismatch.
Decode workloads are memory intensive, while the microbenchmarks in Section~\ref{sec:micro_bench} show that the NPU has only 1 MB L2 cache, compared with 12 MB on one CPU cluster, and its cached read bandwidth is only 178--288 GB/s, versus 480--536 GB/s on the CPU. Once the working set spills into DDR, the NPU's DDR bandwidth reaches only 39.7 GB/s, 1.4--1.5$\times$ lower than the 57--60 GB/s achieved by CPUs and GPUs.
Moreover, decoding operates over a continuously growing context, whereas static-graph NPU execution requires a large context window to be preallocated in advance, introducing avoidable per-token computation and leaving 20.0--23.8\% decode throughput headroom as the context grows from 16 to 4096 tokens.
Together, limited data access speed and static-shape execution prevent the NPU's compute capability from translating into efficient decoding.

\noindent\textbf{(3) Inefficient resource scheduling leaves substantial headroom for optimization.}
Backend-specific energy attribution uncovers critical inefficiencies that remain obscured by coarse, device-level measurements. We find that during NPU-accelerated inference, the host CPU can account for up to 30\% of the {measured energy} due to aggressive polling and synchronization, despite contributing minimally to model computation. By simply fine-tuning RPC polling intervals and NPU sleep latency, energy consumption per token can be reduced by 30.9--37.8\% and 44.6--50.9\%, respectively, typically with negligible throughput degradation. Furthermore, suboptimal thread--core affinity causes up to 35\% performance variance, while fine-grained DVFS (Dynamic Voltage and Frequency Scaling) tuning can recover approximately 50\% of wasted energy on certain backends. These findings underscore that mobile LLM efficiency is not just a kernel-level challenge but a complex scheduling problem requiring both intra-backend calibration and cross-backend coordination.

Collectively, we estimate that coordinating CPU frequency with NPU sleep and polling parameters can reduce energy consumption by up to {53.2\%} on the NPU.

Our contributions are summarized as follows:
\begin{itemize}[leftmargin=*,noitemsep,topsep=2pt]
    \item \textbf{First in-depth NPU measurement:} 
 We present a comprehensive measurement study across five frameworks (\textit{llama.cpp, MNN, GENIE, MLLM, MLC-LLM}) and three hardware backends (CPU, GPU, NPU). By systematically exploring NPU-based deployment, we fill  critical blind spots in prior research and provide a holistic view of the mobile LLM ecosystem.
    \item \textbf{Fine-grained profiling toolkit:} %
    We develop a lightweight, framework-agnostic instrumentation suite that enables fine-grained power and throughput measurements. Unlike prior coarse device-level profiling, our toolkit uncovers hidden energy bottlenecks and execution inefficiencies at the compute-unit level (CPU, GPU, and NPU), providing the precision necessary for deep power analysis.
    \item \textbf{Hidden bottlenecks and optimization opportunities:} %
Through extensive experimentation, we derive actionable insights spanning kernel scalability, activation quantization, and backend coordination. We identify a critical ``phase split'' in efficiency and propose a phase-aware co-design of scheduling and backend selection to guide future mobile LLM optimizations.
    
\end{itemize}

\section{Background}
\label{sec:background}

\subsection{LLM Inference Phases}

LLMs employ transformer architectures~\cite{attention_is_all_you_need} with two distinct inference phases that exhibit fundamentally different computational characteristics:

\textbf{Prefill Phase:} Processes the entire input prompt in parallel to generate the KV cache~\cite{ainslie-etal-2023-gqa_kvcache, 10.1109/ISCA59077.2024.00019}. This phase performs matrix-matrix multiplications (GEMM) with high arithmetic intensity, making it largely compute-bound; performance is therefore determined mainly by the backend's peak FLOPS (floating-point operations per second).

\textbf{Decode Phase:} Generates output tokens autoregressively, one token at a time. This phase performs matrix-vector multiplications (GEMV) as it fetches the entire KV cache for each token generation. Its arithmetic intensity is much lower than that of the prefill phase, making it memory-bound; performance is therefore determined mainly by memory bandwidth.

This phase difference fundamentally shapes backend performance and bottlenecks, as shown in Section~\ref{sec:rq2}.

\begin{figure*}[t]
  \centering
  \begin{minipage}[t]{0.55\textwidth}
    \centering
    \includegraphics[width=\linewidth]{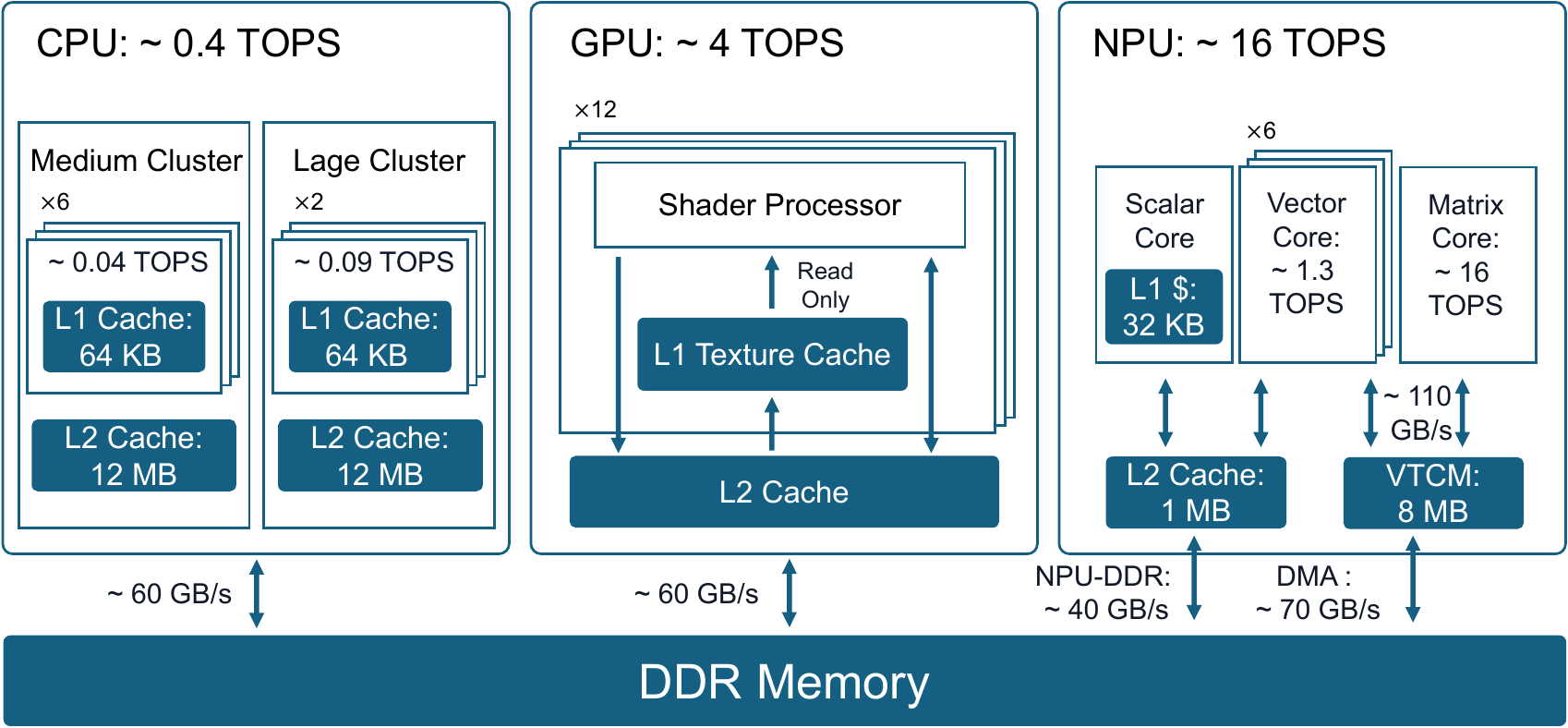}
    \caption{A typical mobile memory and compute-unit architecture. The data are obtained from the hardware microbenchmarks reported in Section~\ref{sec:micro_bench}.}
    \label{fig:arch}
  \end{minipage}
  \hfill
  \begin{minipage}[t]{0.43\textwidth}
    \centering
    \includegraphics[width=\linewidth]{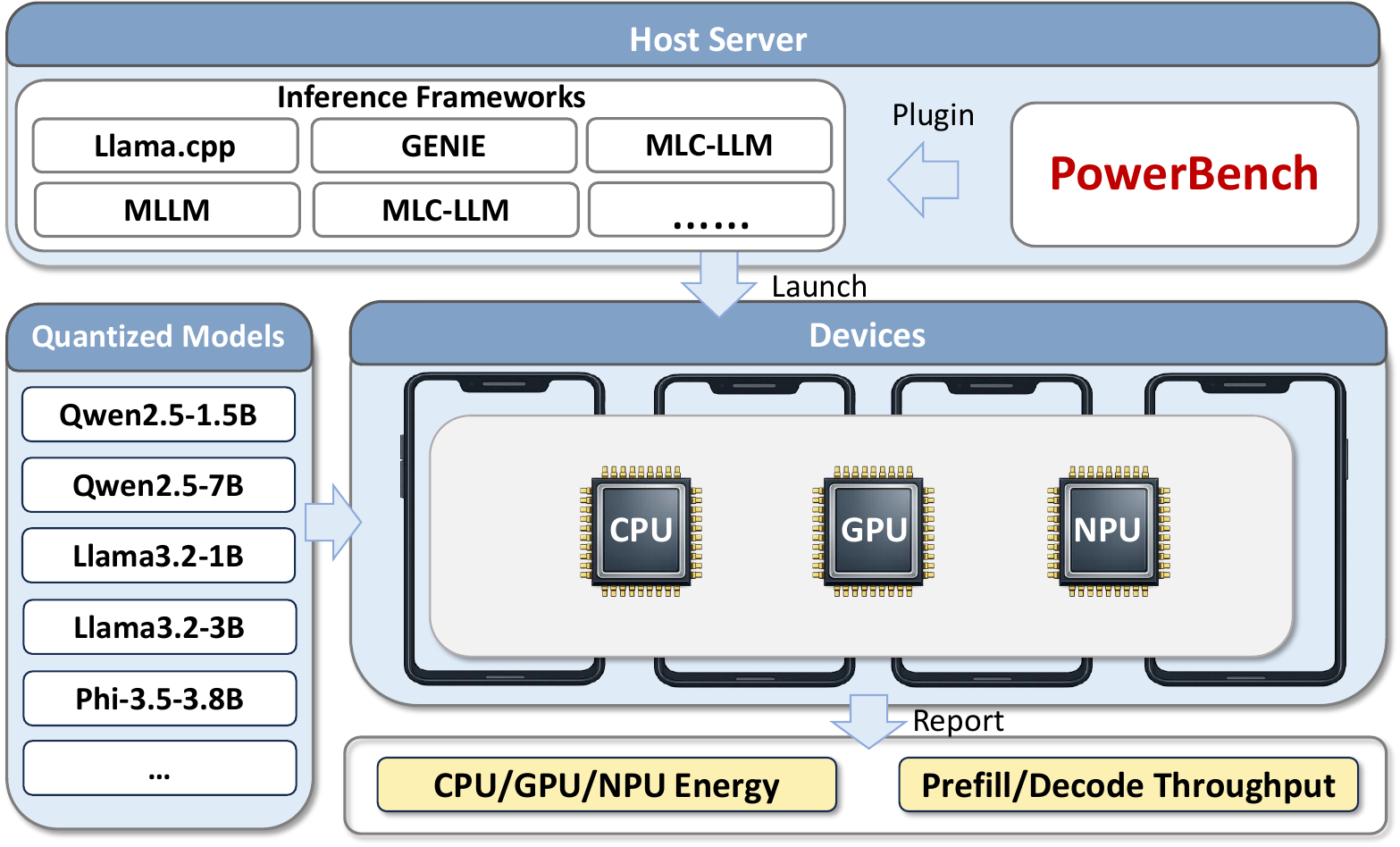}
    \caption{Overview of our evaluation infrastructure.}
    \label{fig:infra-overview}
  \end{minipage}
\end{figure*}

\subsection{Mobile Memory and Heterogeneous Backends}

Modern mobile SoCs integrate three heterogeneous compute units that share the same main DDR memory but differ substantially in peak compute capability, cache hierarchy, and memory-access characteristics. A typical architecture is shown in Figure~\ref{fig:arch}. \textbf{CPUs} consist of ARM big.LITTLE clusters. They offer the greatest flexibility and the lowest control overhead, making them suitable for irregular workloads and latency-sensitive execution. CPUs also have relatively large caches, high effective DDR bandwidth, and low memory-access latency, although their compute capability is limited to approximately 0.5 TOPS (Tera Operations Per Second) under INT32 and FP32. \textbf{GPUs} consist of parallel cores exposed through compute APIs such as OpenCL. Recent mobile GPUs provide around 4 TOPS under FP16, about one order of magnitude higher than CPUs. Their achievable DDR bandwidth is comparable to that of CPUs, but their memory-access latency is higher, making them better suited to highly parallel workloads that can hide this latency. \textbf{NPUs} are dedicated AI accelerators designed for dense tensor computation. Recent NPUs expose over 15 TOPS under INT8 arithmetic, suggesting another order-of-magnitude increase in nominal compute capability. However, this performance usually depends on vendor-specific SDKs and restricted operator support~\cite{noauthor_qualcomm_nodate_qnn,noauthor_hexagon_nodate}, resulting in less flexible execution models. NPUs can fetch DDR data through two paths: the direct NPU-DDR path is slow, while the two-step path can provide high bandwidth but is difficult to fully utilize through programming. We carefully measure latency, bandwidth, and compute capability in Section~\ref{sec:micro_bench} to support an in-depth study of LLM inference speed.

\begin{table}[t]
\centering
\caption{Overview of representative mobile LLM inference frameworks. Symbols denote backend status: \xmark~unsupported; \omark~supported; \cmark~targeted.}
\label{tab:framework_overview}
\scriptsize
\setlength{\tabcolsep}{10pt}
\begin{tabular}{l l l l l l}
\toprule
\textbf{Framework} & \textbf{Maintainer} & \textbf{CPU} & \textbf{GPU} & \textbf{NPU} & \textbf{Version} \\
\midrule
llama.cpp~\cite{llamacpp}   & Community     & \cmark             & \cmark & \cmark~Custom kernel (open source) & \texttt{eadc418} \\
MNN~\cite{jiang2020mnn}         & Alibaba       & \cmark           & \cmark                     & \cmark~QNN kernel (closed source) & \texttt{510ac8f} \\
MLC-LLM~\cite{mlc_llm}     & Community     & \omark                  & \cmark          & \xmark  & \texttt{8f49ea6} \\
MLLM~\cite{mllm}        & Community     & \omark          & \omark                             & \cmark~QNN kernel (closed source) & \texttt{10d3d6a} \\
GENIE~\cite{qualcomm_genie_extensions}       & Qualcomm      & \omark                  & \xmark                         & \cmark~QNN kernel (closed source) & \texttt{2.39.0} \\
\bottomrule
\end{tabular}
\end{table}
\subsection{Mobile Inference Frameworks}

Table~\ref{tab:framework_overview} summarizes representative mobile LLM inference frameworks used in our study and reveals a highly heterogeneous support landscape across different backends.
Among them, llama.cpp has the most active open-source community, MNN benefits from commercial support, MLLM represents an early NPU-oriented LLM inference framework, GENIE provides a Qualcomm vendor-optimized framework, and MLC-LLM offers a compiler-based cross-platform deployment stack.
While CPU and GPU execution are broadly available, NPU support is limited and implemented through two different approaches.
Frameworks such as GENIE, MNN, and MLLM adopt a QNN-based path, delegating execution to Qualcomm's AI Engine Direct and Hexagon software stack and therefore inheriting the same vendor-provided kernels~\cite{noauthor_qualcomm_nodate_qnn,noauthor_hexagon_nodate,qualcomm_genie_extensions}.
In contrast, llama.cpp follows a custom-kernel path, implementing operators at a lower level in open source. This approach provides greater flexibility and control but also demands substantially higher engineering effort.
The table further shows that backend support is not merely a binary capability distinction: frameworks vary in which backends they prioritize and in the maturity of each backend implementation.
This diversity is important for our study because it indicates that performance differences may arise not only from the hardware backends themselves, but also from frameworks' offloading strategies and operator implementation choices, as shown in Section~\ref{sec:rq1}.

\section{Infrastructure and Methodology}

\label{sec:methodology}

\textbf{Macro LLM benchmark:} Figure~\ref{fig:infra-overview} illustrates our end-to-end LLM evaluation infrastructure.
A host machine orchestrates all experiments, including model preparation, deployment, run configuration, and result collection. It communicates with target devices through Android Debug Bridge (ADB).
Before each run, the host follows a protocol that includes thermal conditioning, background isolation, and repeated trials to ensure reproducibility (Section~\ref{subsec:protocol}).
The testbed comprises four smartphones.
On each device, we evaluate five frameworks, namely llama.cpp, MLC-LLM, MLLM, MNN, and GENIE, across the CPU, GPU, and NPU backends.
To collect energy measurements consistently across frameworks, we develop a lightweight instrumentation library \texttt{PowerBench} and plug it into each framework binary (Section~\ref{subsec:energy}).
For throughput measurements, the same harness standardizes benchmark settings across frameworks, including token size, runtime, and repeated runs (Section~\ref{subsec:throughput}).
The evaluated models include Llama~3.2-1B, Llama~3.2-3B, Qwen~2.5-1.5B, Qwen~2.5-7B, and Phi~3.5-mini (3.8B).
{All models use 4-bit integer weight quantization (W4). Activation tensor formats are FP16 or FP32, except in the W4A16 NPU configurations in Table~\ref{tab:throughput_across_device_model_framework_backend}, which use 16-bit integers. KV caches use FP16, except for GENIE on CPU (FP32) and the W4A16 NPU configurations (8-bit integers). We evaluate model quality using PPL (Section~\ref{subsec:quality}).}
Overall, this controlled benchmark space spans more than 400 configurations across models, devices, frameworks, backends, and inference stages.
The results are streamed back to the host for offline analysis.

\textbf{Micro hardware benchmark:} To understand the underlying hardware characteristics, we conduct a set of isolated microbenchmarks on the memory and compute units. For memory, we measure both latency and bandwidth. Latency is measured using pointer-chasing workloads over increasing working-set sizes. This sweep exposes transitions across the L1 cache, L2 cache, and DRAM. Bandwidth is measured using streaming read kernels, with large buffers for DRAM bandwidth and small or explicitly placed buffers for cache and scratchpad throughput. For compute capability, we use NEON FMA on the CPU; OpenCL kernels that repeatedly execute multiply-add operations on the GPU; and HVX (Hexagon Vector eXtensions) \texttt{mpyacc} and \texttt{vrmpyacc} instructions together with HMX (Hexagon Matrix eXtensions) matrix multiplication on the Hexagon NPU. We report all results using a unified convention in which one FMA or MAC counts as two operations. These microbenchmarks characterize near-peak hardware limits.

\begin{table}[t]
\caption{Devices used in our experiments.}
\centering
\scriptsize
\renewcommand{\arraystretch}{1.12}
\setlength{\tabcolsep}{4pt}
\begin{tabular}{l c c c c c c}
\toprule
\textbf{Device} & \textbf{Year} & \textbf{SoC} & \textbf{CPU} & \textbf{GPU} & \textbf{NPU} & \textbf{DRAM} \\
\midrule
Xiaomi 17  & 2025 & SM8850 & 2$\times$4.6 GHz + 6$\times$3.62 GHz & Adreno 840, 1.20 GHz & V81 & 16 GB \\
OnePlus 15 & 2025 & SM8850 & 2$\times$4.6 GHz + 6$\times$3.62 GHz & Adreno 840, 1.20 GHz & V81 & 16 GB \\
Xiaomi 15  & 2024 & SM8750      & 2$\times$4.32 GHz + 6$\times$3.53 GHz & Adreno 830, 1.10 GHz & V79 & 16 GB \\
Xiaomi 14  & 2023 & SM8650 &
1$\times$3.3 GHz + 3$\times$3.15 GHz + 2$\times$2.96 GHz + 2$\times$2.27 GHz
& Adreno 750, 1.00 GHz & V75 & 16 GB \\
\bottomrule
\end{tabular}
\label{tab:devices}
\end{table}

\subsection{Hardware Platform}
\label{subsec:devices}
We conduct experiments on four smartphones spanning three generations of Qualcomm Snapdragon platforms, as summarized in Table~\ref{tab:devices}.
The testbed covers SoCs from 2023 to 2025, including SM8650, SM8750, and SM8850, capturing both generational evolution and device-level variation.
Across these devices, CPU and GPU frequencies increase steadily across generations, and the NPU evolves from V75 to V81. All phones provide 16~GB DRAM, offering a consistent memory-capacity baseline.
We use Qualcomm platforms because they are common in prior mobile LLM studies, especially those involving NPU execution~\cite{xuFastOndeviceLLM2024,chung_scaling_2024,xuEdgeLLMFastOnDevice2025}, and because their NPU stack currently has the broadest support across mainstream frameworks, enabling controlled evaluation across CPU, GPU, and NPU backends.
{We selected Xiaomi devices because they cover multiple SoC generations and are easy to root, enabling power and energy measurements and fixed-frequency experiments. We additionally included the OnePlus 15 to compare different devices using the same SM8850 SoC.}

\subsection{Throughput Profiling}
\label{subsec:throughput}

We measure prefill and decode throughput. We also measure layer- and operator-level execution latency.
Because different frameworks expose different benchmark and runtime interfaces, we build a benchmark tool for each framework and standardize the measurement protocol by controlling token size, benchmark runtime, and the number of repeated runs.
This design keeps the workload scale and measurement duration comparable.
Prefill throughput is computed as the prompt length divided by the elapsed time from inference start to the emission of the first output token.
Decode throughput is computed as the number of generated tokens divided by the elapsed time between the first and the last emitted output tokens.
To analyze the sources of performance differences across frameworks, we benchmark representative layers and operators covering the dominant LLM workloads, including attention and FFN layers, prefill GEMM operators, and decode GEMV operators.
For each operator, we construct isolated micro-benchmarks using tensor shapes that match those encountered in real inference.

\subsection{Energy Profiling}
\label{subsec:energy}

We use four per-token energy metrics (\textmu J/token): SoC, CPU, GPU, and NPU. Here, SoC energy denotes total chip energy, including CPU, GPU, NPU, and other on-chip components, enabling efficiency analysis at multiple granularities.
To attribute energy consumption to individual compute units, we develop \texttt{PowerBench}, a lightweight C++ instrumentation library built on top of the Qualcomm Power Telemetry (QPT) driver stack~\cite{oneplus_qpt_kernel}.
QPT exposes cumulative energy counters for power management integrated circuit (PMIC)-regulated power zones through the Linux \texttt{powercap} sysfs interface, covering the whole SoC as well as rails for the CPU clusters, GPU, and NPU.
Unlike coarse device-level approaches such as Android \texttt{dumpsys batterystats} or external power monitors~\cite{huangMNNAECSEnergyOptimization2025,laskaridisMELTingPointMobile2024}, \texttt{PowerBench} automatically discovers available power zones, maps them to logical compute units, and records energy deltas through \texttt{start()}/\texttt{stop()} calls.
Total energy is derived from counter differences, while a background sampling thread records temporal power traces at configurable intervals for auxiliary analysis.
The library is header-only and can be integrated into existing C++ benchmarks with minimal code changes.
We use it uniformly across frameworks to ensure measurement consistency.
This backend-level granularity allows us to identify and optimize energy waste that coarse device-level measurements cannot isolate, as shown in Section~\ref{sec:rq3}.

\subsection{{Quality Profiling}}
\label{subsec:quality}

{We assess model output quality using perplexity (PPL), the exponential of mean next-token cross entropy; lower values indicate better text prediction~\cite{bengio2003neuralprobabilistic}. All PPL measurements use instruction-tuned models and the same fixed 10\% line sample of the WikiText-2 test split~\cite{merity2017pointersentinel}. We evaluate the sampled text with 128-token windows and an 84-token stride, scoring each next-token target once. Each scored token receives equal weight.}

\subsection{Experimental Protocol}
\label{subsec:protocol}

To ensure measurement reliability and reproducibility, we follow a strict experimental protocol. {Before each experiment, devices are cooled to an initial battery temperature below 28$^\circ$C to limit thermal throttling and residual-heat effects. This threshold is an upper bound on the initial temperature. The minimum battery temperature recorded during inference was 15.8$^\circ$C.} We further disable background services, notifications, and network connectivity. Devices run in airplane mode with the screen turned off, and only the inference process is active during measurement. {By default, we pin the inference process to the CPU cores specified by the affinity mask \texttt{0xe0}}. Each configuration includes one warm-up run followed by at least three recorded trials, and we report the mean across repeated runs unless otherwise stated.

\section{Macro LLM Benchmark and Research Questions}
\label{sec:overview}

\begin{table*}[t]
\centering
\caption{Throughput and energy across devices, models, frameworks, and backends for 256-token workloads. Perplexity (PPL) is measured on the WikiText-2 test split.}
\label{tab:throughput_across_device_model_framework_backend}
\scriptsize
\setlength{\tabcolsep}{1.2pt}
\renewcommand{\arraystretch}{1}
\resizebox{\textwidth}{!}{%
\begin{tabular}{@{}c c c c c *{12}{c}@{}}
\toprule
\multirow{4}{*}{Model} & \multirow{4}{*}{Backend} & \multirow{4}{*}{Framework} & \multirow{4}{*}{Quan.} & \multirow{4}{*}{PPL~$\downarrow$}
& \multicolumn{6}{c}{Prefill} & \multicolumn{6}{c}{Decode} \\
\cmidrule(lr){6-11}\cmidrule(lr){12-17}
& & & &
& \multicolumn{4}{c}{Throughput (tok/s)} & \multicolumn{2}{c}{Energy ($\mu$J/tok)}
& \multicolumn{4}{c}{Throughput (tok/s)} & \multicolumn{2}{c}{Energy ($\mu$J/tok)} \\
\cmidrule(lr){6-9}\cmidrule(lr){10-11}\cmidrule(lr){12-15}\cmidrule(lr){16-17}
& & & &
& \makecell{Xiaomi\\17} & \makecell{OnePlus\\15} & \makecell{Xiaomi\\15} & \makecell{Xiaomi\\14} & \makecell{Xiaomi\\17} & \makecell{OnePlus\\15}
& \makecell{Xiaomi\\17} & \makecell{OnePlus\\15} & \makecell{Xiaomi\\15} & \makecell{Xiaomi\\14} & \makecell{Xiaomi\\17} & \makecell{OnePlus\\15} \\
\midrule

\multirow{10}{*}{Qwen2.5-1.5B}
 & CPU & GENIE & w4 & {20.9} & 135.6 & 123.6 & 129.2 & 98.9 & 3.6e4 & 3.3e4 & 31.3 & 27.4 & 31.1 & 20.9 & 8.0e4 & 9.5e4 \\
 & CPU & llama.cpp & w4 & {18.1} & {417.4} & 299.9 & 329.2 & 157.2 & 3.3e4 & 1.5e4 & \underline{\textbf{51.6}} & \underline{\textbf{55.7}} & \underline{\textbf{53.9}} & 34.5 & 1.2e5 & 9.7e4 \\
 & CPU & MNN & w4 & {19.7} & {187.8} & 228.3 & 259.4 & 259.5 & 2.3e4 & 1.5e4 & {42.3} & 47.5 & 49.5 & \underline{\textbf{45.7}} & 1.3e5 & 9.2e4 \\
 & GPU & llama.cpp & w4 & {18.0} & 569.7 & 754.8 & 680.9 & 365.7 & 1.0e4 & 1.1e4 & {38.9} & 50.3 & 48.6 & 31.8 & 1.1e5 & 1.4e5 \\
 & GPU & MNN & w4 & {18.8} & {586.6} & 434.1 & 406.1 & 272.5 & 8.6e3 & 1.7e4 & {33.4} & 12.3 & 45.8 & 26.2 & 9.4e4 & 9.9e4 \\
 & GPU & MLC-LLM & w4 & -- & 24.3 & 42.4 & 45.9 & 141.8 & 5.7e4 & 5.0e4 & 24.6 & 19.6 & 29.9 & 16.8 & 1.3e5 & 1.3e5 \\
 & NPU & GENIE & w4 & {19.3} & {\underline{\textbf{1625.2}}} & \underline{\textbf{1463.7}} & \underline{\textbf{1219.2}} & -- & 6.8e3 & 5.6e3 & 23.1 & 23.0 & 20.7 & -- & 4.3e5 & 3.2e5 \\
 & NPU & llama.cpp & w4 & {18.2} & 107.6 & 115.1 & 82.3 & 51.4 & 8.2e4 & 7.6e4 & 31.9 & 33.3 & 18.3 & 13.8 & 2.0e5 & 1.6e5 \\
 & NPU & MNN & w4 & {18.7} & {818.1} & 700.9 & 694.4 & \underline{\textbf{593.8}} & 8.3e3 & 8.1e3 & {14.7} & 10.3 & 16.9 & 14.8 & 2.1e5 & 2.5e5 \\
 & NPU & MLLM & w4a16 & {23.0} & 904.5 & 966.8 & 814.6 & -- & \underline{\textbf{3.2e3}} & \underline{\textbf{4.1e3}} & 34.0 & 34.3 & 32.7 & -- & \underline{\textbf{7.5e4}} & \underline{\textbf{8.7e4}} \\
\midrule

\multirow{8}{*}{Qwen2.5-7B}
 & CPU & GENIE & w4 & {15.4} & 27.8 & 26.8 & 29.0 & 18.9 & 1.7e5 & 1.5e5 & 7.5 & 6.7 & 7.1 & 5.4 & 3.5e5 & 4.2e5 \\
 & CPU & llama.cpp & w4 & {13.6} & 85.4 & 58.6 & 55.1 & 34.1 & 1.1e5 & 7.9e4 & 12.6 & 14.3 & 13.2 & 8.7 & 5.0e5 & 4.4e5 \\
 & CPU & MNN & w4 & {15.3} & 80.3 & 30.0 & 58.4 & 50.8 & 9.8e4 & 1.0e5 & 9.6 & 12.2 & 12.2 & 7.2 & 4.6e5 & \underline{\textbf{3.9e5}} \\
 & GPU & llama.cpp & w4 & {13.9} & 137.5 & 171.6 & 141.8 & \underline{\textbf{84.6}} & 4.0e4 & 6.5e4 & 10.2 & 12.2 & 11.7 & 5.4 & 3.2e5 & 5.7e5 \\
 & GPU & MNN & w4 & {15.2} & 78.4 & 102.1 & 86.2 & 62.4 & 4.1e4 & 9.1e4 & 8.9 & 10.6 & 12.7 & \underline{\textbf{11.7}} & \underline{\textbf{3.1e5}} & 5.1e5 \\
 & GPU & MLC-LLM & w4 & -- & -- & -- & -- & 29.7 & -- & -- & 10.0 & 13.4 & 11.2 & 8.0 & 5.9e5 & 8.8e5 \\
 & NPU & GENIE & w4a16 & {16.3} & \underline{\textbf{859.1}} & \underline{\textbf{1411.1}} & \underline{\textbf{1020.9}} & -- & \underline{\textbf{9.4e3}} & \underline{\textbf{6.9e3}} & \underline{\textbf{14.6}} & \underline{\textbf{16.8}} & \underline{\textbf{13.6}} & -- & 4.0e5 & 4.0e5 \\
 & NPU & llama.cpp & w4 & {$>1000$} & 34.8 & 35.1 & 27.0 & 13.0 & 1.6e5 & 6.0e5 & 9.1 & 9.2 & 7.7 & 6.4 & 6.9e5 & 6.0e5 \\
\midrule

\multirow{10}{*}{Llama3.2-1B}
 & CPU & GENIE & w4 & {30.8} & 182.2 & 167.4 & 177.3 & 134.4 & 2.8e4 & 2.3e4 & 39.9 & 35.6 & 39.3 & 28.1 & 6.3e4 & 7.5e4 \\
 & CPU & llama.cpp & w4 & {26.8} & 326.8 & 287.6 & 386.3 & 166.3 & 1.1e4 & 1.6e4 & \underline{\textbf{76.1}} & \underline{\textbf{72.2}} & \underline{\textbf{70.1}} & 41.9 & 1.4e5 & 7.4e4 \\
 & CPU & MNN & w4 & {27.6} & 29.8 & 184.8 & 324.8 & 321.9 & 8.7e3 & 1.5e4 & 3.2 & 61.6 & 66.6 & \underline{\textbf{55.4}} & 7.9e4 & \underline{\textbf{6.6e4}} \\
 & GPU & llama.cpp & w4 & {26.8} & 744.0 & 986.8 & 862.1 & 450.1 & 6.3e3 & 5.4e3 & 46.1 & 64.4 & 57.1 & 22.5 & 1.4e5 & 1.1e5 \\
 & GPU & MNN & w4 & {27.1} & 692.1 & 669.8 & 534.4 & 371.3 & 6.9e3 & 1.3e4 & 16.8 & 21.6 & 60.7 & 44.4 & \underline{\textbf{5.5e4}} & 7.9e4 \\
 & GPU & MLC-LLM & w4 & -- & -- & -- & -- & 189.3 & -- & -- & 31.6 & 19.9 & 40.1 & 20.1 & 1.0e5 & 9.3e4 \\
 & NPU & GENIE & w4 & {31.4} & 1887.1 & 2316.1 & \underline{\textbf{1747.9}} & -- & 4.2e3 & 4.4e3 & 25.9 & 26.3 & 23.6 & -- & 2.9e5 & 3.0e5 \\
 & NPU & llama.cpp & w4 & {26.8} & 132.0 & 141.3 & 93.5 & 42.3 & 7.6e4 & 4.6e4 & 41.3 & 41.6 & 21.0 & 13.8 & 2.4e5 & 1.3e5 \\
 & NPU & MNN & w4 & {27.1} & 1215.6 & 228.9 & 1043.9 & 850.7 & 5.7e3 & 6.1e3 & 13.8 & 13.8 & 21.1 & 18.6 & 1.8e5 & 1.9e5 \\
 & NPU & MLLM & w4a16 & {46.4} & \underline{\textbf{2163.2}} & \underline{\textbf{2322.2}} & 1716.0 & \underline{\textbf{1624.4}} & \underline{\textbf{2.8e3}} & \underline{\textbf{3.2e3}} & 58.4 & 62.5 & 50.9 & 51.3 & 6.3e4 & 8.6e4 \\
\midrule

\multirow{10}{*}{Llama3.2-3B}
 & CPU & GENIE & w4 & {21.7} & 39.7 & 59.2 & 62.1 & 45.7 & 6.2e4 & 6.8e4 & 15.9 & 14.0 & 15.7 & 12.0 & \underline{\textbf{1.7e5}} & 2.0e5 \\
 & CPU & llama.cpp & w4 & {20.1} & 119.7 & 109.8 & 114.8 & 58.6 & 5.4e4 & 4.1e4 & \underline{\textbf{29.9}} & \underline{\textbf{30.5}} & \underline{\textbf{27.1}} & 17.1 & 2.9e5 & 2.1e5 \\
 & CPU & MNN & w4 & {21.9} & 123.4 & 66.1 & 90.8 & 113.1 & 6.8e4 & 4.4e4 & 12.4 & 24.9 & 24.0 & \underline{\textbf{23.7}} & 2.1e5 & \underline{\textbf{1.8e5}} \\
 & GPU & llama.cpp & w4 & {20.2} & 224.8 & 357.6 & 306.6 & 161.1 & 2.2e4 & 2.1e4 & 20.0 & 25.0 & 25.2 & 11.1 & 1.8e5 & 2.5e5 \\
 & GPU & MNN & w4 & {21.8} & 201.2 & 248.7 & 205.3 & 145.3 & 2.2e4 & 3.5e4 & 24.4 & 12.1 & 25.4 & 17.1 & 2.0e5 & 2.0e5 \\
 & GPU & MLC-LLM & w4 & -- & -- & -- & -- & 67.8 & -- & -- & 19.8 & 20.9 & 21.2 & 13.3 & 2.8e5 & 3.9e5 \\
 & NPU & GENIE & w4 & {22.6} & 615.7 & 815.3 & \underline{\textbf{656.3}} & -- & 1.3e4 & 1.1e4 & 8.9 & 12.2 & 10.8 & -- & 7.0e5 & 7.1e5 \\
 & NPU & llama.cpp & w4 & {20.2} & 48.1 & 43.0 & 34.1 & 15.3 & 1.6e5 & 1.2e5 & 19.1 & 19.3 & 10.1 & 3.9 & 4.4e5 & 2.9e5 \\
 & NPU & MNN & w4 & {21.7} & 533.5 & 97.2 & 454.5 & 440.1 & 1.3e4 & 1.4e4 & 6.0 & 10.5 & 8.2 & 7.5 & 4.3e5 & 7.3e5 \\
 & NPU & MLLM & w4a16 & {43.5} & \underline{\textbf{929.3}} & \underline{\textbf{886.2}} & -- & \underline{\textbf{674.8}} & \underline{\textbf{7.4e3}} & \underline{\textbf{8.3e3}} & -- & -- & -- & -- & -- & -- \\
\midrule

\multirow{6}{*}{Phi-3.5-mini}
 & CPU & GENIE & w4 & {13.0} & -- & -- & 52.0 & 37.5 & -- & -- & -- & -- & 13.6 & 9.1 & -- & -- \\
 & CPU & llama.cpp & w4 & {11.4} & 97.3 & 83.3 & 85.6 & 46.8 & 8.2e4 & 4.5e4 & \underline{\textbf{24.1}} & \underline{\textbf{24.9}} & 22.4 & \underline{\textbf{15.4}} & 3.3e5 & \underline{\textbf{2.6e5}} \\
 & GPU & llama.cpp & w4 & {11.4} & \underline{\textbf{216.0}} & \underline{\textbf{281.3}} & 228.8 & 118.1 & \underline{\textbf{2.0e4}} & \underline{\textbf{3.5e4}} & 19.6 & 24.1 & \underline{\textbf{22.7}} & 10.2 & \underline{\textbf{1.9e5}} & 2.8e5 \\
 & GPU & MLC-LLM & w4 & -- & -- & -- & -- & 53.1 & -- & -- & 19.3 & 20.9 & 20.1 & 14.7 & 3.3e5 & 4.0e5 \\
 & NPU & GENIE & w4a16 & {12.6} & -- & -- & \underline{\textbf{1039.4}} & \underline{\textbf{746.8}} & -- & -- & -- & -- & 12.6 & 9.1 & -- & -- \\
 & NPU & llama.cpp & w4 & {$>1000$} & 40.3 & 44.1 & 26.3 & 18.5 & 1.7e5 & 1.3e5 & 2.7 & 3.0 & 4.6 & 1.4 & 1.6e6 & 1.3e6 \\
\bottomrule
\end{tabular}
}
\end{table*}

{We first conduct a benchmark across devices, models, frameworks, and backends to establish a holistic view of the current performance landscape. All frameworks are evaluated under their default high-performance settings, with 4-bit weight quantization and no activation quantization unless otherwise noted. This benchmark provides a unified view of throughput and energy for 256-token prefill and decode workloads. The results are shown in Table~\ref{tab:throughput_across_device_model_framework_backend}.}
{Table~\ref{tab:throughput_repeatability} reports throughput repeatability and the maximum battery temperature recorded for each configuration.} {Our experiments target peak inference performance under controlled thermal conditions, and no thermal throttling occurred in these experiments.}

\begin{table}[t]
{
\centering
\caption{{Throughput repeatability and maximum battery temperature for Qwen2.5-1.5B on Xiaomi 17. Throughput is reported as mean $\pm$ standard deviation (SD) in tok/s; the coefficient of variation (CV) is SD divided by the mean.}}
\label{tab:throughput_repeatability}
\scriptsize
\newcommand{\stabilitysd}[1]{{\fontsize{6}{7}\selectfont$\pm #1$}}
\setlength{\tabcolsep}{1.2pt}
\renewcommand{\arraystretch}{1.08}
\resizebox{\linewidth}{!}{%
\begin{tabular}{@{}l l *{10}{r}@{}}
\toprule
\multirow{2}{*}{Phase} & \multirow{2}{*}{Metric}
& \multicolumn{3}{c}{CPU} & \multicolumn{3}{c}{GPU} & \multicolumn{4}{c}{NPU} \\
\cmidrule(lr){3-5}\cmidrule(lr){6-8}\cmidrule(lr){9-12}
& & GENIE & llama.cpp & MNN & llama.cpp & MNN & MLC-LLM & GENIE & llama.cpp & MNN & MLLM \\
\midrule
\multirow{3}{*}{Prefill} & Mean $\pm$ SD & $135.6$\,\stabilitysd{0.15} & $417.4$\,\stabilitysd{24.02} & $187.8$\,\stabilitysd{10.89} & $569.7$\,\stabilitysd{0.60} & $586.6$\,\stabilitysd{0.42} & $24.3$\,\stabilitysd{0.03} & $1625.2$\,\stabilitysd{6.15} & $107.6$\,\stabilitysd{0.37} & $818.1$\,\stabilitysd{18.55} & $904.5$\,\stabilitysd{4.14} \\
 & CV (\%) & 0.11 & 5.76 & 5.80 & 0.10 & 0.07 & 0.13 & 0.38 & 0.34 & 2.27 & 0.46 \\
 & Max. temp. ($^\circ$C) & 18.8 & 21.2 & 18.3 & 25.0 & 23.8 & 23.6 & 20.0 & 24.0 & 23.8 & 21.9 \\
\midrule
\multirow{3}{*}{Decode} & Mean $\pm$ SD & $31.3$\,\stabilitysd{0.02} & $51.6$\,\stabilitysd{0.21} & $42.3$\,\stabilitysd{1.94} & $38.9$\,\stabilitysd{0.11} & $33.4$\,\stabilitysd{0.34} & $24.6$\,\stabilitysd{0.05} & $23.1$\,\stabilitysd{0.31} & $31.9$\,\stabilitysd{0.33} & $14.7$\,\stabilitysd{0.15} & $34.0$\,\stabilitysd{0.79} \\
 & CV (\%) & 0.06 & 0.40 & 4.58 & 0.28 & 1.02 & 0.21 & 1.34 & 1.03 & 1.05 & 2.34 \\
 & Max. temp. ($^\circ$C) & 20.0 & 23.4 & 23.9 & 25.4 & 24.2 & 24.9 & 24.1 & 25.0 & 29.2 & 23.6 \\
\bottomrule
\end{tabular}%
}
}
\end{table}

\textbf{Overall trends.}
First, prefill throughput is usually highest on NPU when a mature NPU path is available; many GENIE and MLLM entries exceed 1,000 tokens/s, while CPU and GPU entries are lower. Second, decode throughput shifts toward CPU and GPU: for small and medium models, CPU and GPU often remain in the 20--70 tokens/s range, while NPU decode frequently falls behind. Third, model size and device generation behave largely as expected, as larger models reduce throughput and increase per-token energy.

\textbf{Notable observations.}
Beyond these expected trends, the table also reveals several observations that cannot be explained by model size or backend class alone. (1) Framework gaps are strongly amplified on NPUs. On the same OnePlus 15 device and Qwen2.5-1.5B model, GENIE reaches 1463.7 tokens/s in NPU prefill, while MNN reaches 700.9 tokens/s and llama.cpp reaches only 115.1 tokens/s; the same model on CPU or GPU shows much smaller framework gaps. (2) The best backend flips sharply between prefilling and decoding. For example, on Qwen2.5-1.5B decode on OnePlus 15, llama.cpp on CPU reaches 55.7 tokens/s, compared with 23.0 tokens/s for GENIE on NPU, indicating a phase-dependent backend split rather than a single universally best accelerator. (3) NPUs are widely marketed as highly energy-efficient AI accelerators, yet their measured decode energy is often surprisingly high. On OnePlus 15 with Qwen2.5-1.5B decode, GENIE on NPU consumes $3.2\times 10^{5}$~$\mu$J/token, compared with $9.7\times 10^{4}$~$\mu$J/token for llama.cpp on CPU and $1.4\times 10^{5}$~$\mu$J/token for llama.cpp on GPU. The same pattern appears on Xiaomi 17, where GENIE's NPU decode consumes $4.3\times 10^{5}$~$\mu$J/token, far above $1.2\times 10^{5}$~$\mu$J/token on CPU and $1.1\times 10^{5}$~$\mu$J/token on GPU. Thus, despite the NPU's reputation for energy efficiency, its end-to-end decode energy can be roughly 2--4$\times$ higher than CPU/GPU alternatives. These anomalies suggest that the efficiency problem spans framework design, backend architecture, and resource scheduling.

These observations motivate three research questions:
\begin{itemize}[leftmargin=*,noitemsep,topsep=2pt]
    \item \textbf{RQ1: Framework gaps.} What causes the large performance gaps across NPU frameworks?
    \item \textbf{RQ2: Backend optimization.} Why does backend performance ranking reverse between prefilling and decoding?
    \item \textbf{RQ3: Resource scheduling.} How effective is resource scheduling in reducing unnecessary energy overhead?
\end{itemize}

\section{Micro Hardware Benchmark}
\label{sec:micro_bench}

\begin{figure*}[t]
  \centering

  \begin{subfigure}[t]{0.32\textwidth}
    \centering
    \includegraphics[width=\linewidth]{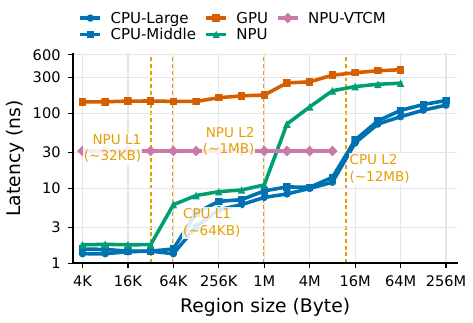}
    \caption{Memory latency vs. size.}
    \label{fig:memory_latency_size}
  \end{subfigure}
  \hfill
  \begin{subfigure}[t]{0.32\textwidth}
    \centering
    \includegraphics[width=\linewidth]{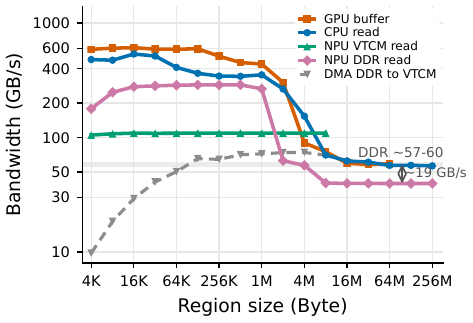}
    \caption{Memory bandwidth vs. size.}
    \label{fig:memory_bandwidth_size}
  \end{subfigure}
  \hfill
  \begin{subfigure}[t]{0.32\textwidth}
    \centering
    \includegraphics[width=\linewidth]{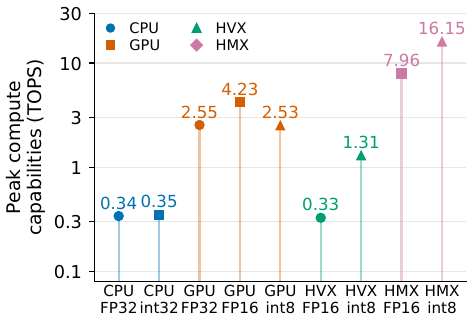}
    \caption{Peak compute capabilities.}
    \label{fig:compute_tops}
  \end{subfigure}
  \captionsetup{skip=1pt}
  \caption{Microbenchmark results for hierarchical memory latency and bandwidth, and peak CPU, GPU, and NPU compute capabilities across precisions on Xiaomi 15.}
  \label{fig:micro_benchmarks}
\end{figure*}

Beyond benchmarking end-to-end LLM inference, we further conduct micro hardware benchmarks on Xiaomi 15 to characterize the underlying memory and compute resources. Figure~\ref{fig:micro_benchmarks} reports three measurements: read latency across increasing region sizes, memory bandwidth across cache and DDR regimes, and peak compute capability across CPU, GPU, and NPU execution units. These measurements help explain why the same LLM workload can favor different backends across prefill and decode.

Figure~\ref{fig:memory_latency_size} first shows that dependent read latency follows different regimes across backends. For small regions, CPU and cached NPU accesses are both near 1--2~ns, while GPU latency is already about 141--146~ns. As the region grows, the CPU remains below 15~ns up to 8~MB and then rises to 40~ns at 16~MB and 128--148~ns at 256~MB. In contrast, NPU DDR latency rises much earlier, from 11~ns at 1~MB to 72~ns at 2~MB and 251~ns at 64~MB. The GPU remains the highest-latency path, reaching about 380~ns at 64~MB. The same sweep reveals approximate capacity boundaries: the CPU has an L1-sized region around 64~KB and a larger L2-scale region around 12~MB, whereas the NPU shows an L1-sized region around 32~KB and an L2-scale region around 1~MB. Thus, the CPU provides a much larger low-latency cache window for irregular or reuse-heavy workloads, while GPU and NPU designs rely more heavily on parallelism and explicit data movement to tolerate memory latency.

Figure~\ref{fig:memory_bandwidth_size} then shows a different ordering for bandwidth. When accesses hit on-chip cache, GPU $>$ CPU $>$ NPU: the GPU buffer path reaches roughly 590--610~GB/s for small regions, the 8-core CPU read path reaches about 480--536~GB/s, and cached NPU DDR reads reach roughly 178--288~GB/s before exceeding the NPU cache regime. Once the region reaches DDR-sized regimes, CPU and GPU bandwidth converge around 57--60~GB/s. The NPU DDR read path reaches about 39.7~GB/s with multi-worker HVX reads, while a single HVX worker reaches only about 11.6~GB/s for large regions. CPU/GPU DDR bandwidth is therefore about 1.4--1.5$\times$ higher than the measured aggregate NPU DDR read path. This gap matters for LLM decode, where each generated token repeatedly streams weights and KV-cache data with limited arithmetic intensity.

The NPU also has a second memory path: data can be copied from DDR into VTCM by DMA and then consumed by the NPU from VTCM. DDR-to-VTCM DMA reaches 74 GB/s, while VTCM-to-NPU reads reach about 109 GB/s. A naive serial implementation has an effective bandwidth of roughly $\left(1/74 + 1/109\right)^{-1} \approx 44$~GB/s, because every byte must first be moved into VTCM and then read again by the compute unit. With careful double buffering, the theoretical streaming limit improves to $\min(74,109) \approx 74$~GB/s by overlapping DMA transfers with NPU computation. However, achieving this bound requires explicit VTCM tiling, asynchronous DMA scheduling, and per-kernel double-buffer management, which substantially increase engineering complexity. Moreover, DMA has high startup latency, so only kernels larger than 1~MB can saturate the DMA path.

Figure~\ref{fig:compute_tops} shows that low-precision peak compute strongly favors the NPU matrix unit. HMX reaches about 16.15~TOPS for INT8 tile microbenchmarks and 7.96~TFLOPS for FP16, while the GPU reaches 4.23~TFLOPS for FP16 and about 2.53~TOPS for INT8. The HVX vector unit reaches about 1.31~TOPS for INT8, and multicore CPU compute remains below 0.4~TOPS. This explains why NPU backends can dominate prefill when large GEMM operators expose enough parallelism to use HMX efficiently.

\section{RQ1: Framework Gaps}
\label{sec:rq1}

Mobile inference frameworks share a common pipeline, including graph conversion, quantization, backend selection, operator offloading, and runtime optimization. The benchmark overview in Section~\ref{sec:overview} shows that framework gaps are particularly large on NPUs, especially in prefill. We find three major causes behind this amplification: offloading strategy, operator scaling, and activation quantization.

\subsection{Finding 1: Substantial Offloading Overhead}
\label{sec:rq1_finding1}

\begin{table}[t]
\centering
\caption{Workload breakdown for Qwen2.5-1.5B: GENIE (compute cycles) vs. MNN (end-to-end latency).}
\label{tab:qnn_mnn_offloading_breakdown}
\scriptsize
\setlength{\tabcolsep}{6pt}
\begin{tabular}{l l r}
\toprule
\textbf{Framework} & \textbf{Component} & \textbf{Latency Share} \\
\midrule
\multirow{2}{*}{\textbf{GENIE}} & NPU Attention & 44.79\% \\
& NPU others & 55.21\% \\
\midrule
\multirow{2}{*}{\textbf{MNN}} & CPU Attention & 1.06\% \\
& NPU Plugins & 98.94\% \\
\bottomrule
\end{tabular}
\end{table}

Both GENIE and MNN use Qualcomm QNN for NPU acceleration, yet their performance differs sharply. GENIE achieves nearly 2$\times$ the throughput of MNN. We find that this gap is mainly caused by different offloading granularity. We dump the computation graph and find that GENIE executes the full graph on the NPU, whereas MNN keeps self-attention on the CPU and offloads the feed-forward and other modules to the NPU. 

Table~\ref{tab:qnn_mnn_offloading_breakdown} shows a clear divergence between compute share and runtime share. In GENIE, non-attention modules account for $55.21\%$ of true NPU compute cycles, but in MNN, NPU plugins occupy $98.94\%$ of end-to-end latency. This indicates that MNN's plugin time includes substantial overhead beyond computation, such as CPU--NPU remote process calls and memory layout conversion, which significantly reduce NPU efficiency.

\subsection{Finding 2: Hardware-Underutilizing Operators}
\begin{figure}[t]
  \centering
  \includegraphics[width=0.55\linewidth]{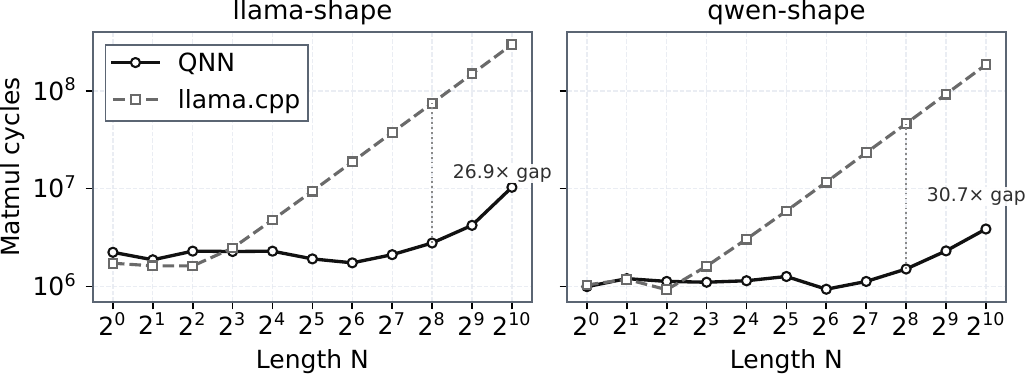}
  \caption{Comparison of matmul cycles between the QNN implementation used by GENIE and llama.cpp.}
  \label{fig:matmul_cycles}
\end{figure}

On Qwen2.5-1.5B with the OnePlus 15 NPU backend, GENIE reaches 1463.7 tokens/s in prefill, while llama.cpp reaches only 115.1 tokens/s, over a 10$\times$ gap. In decode, however, their throughputs are nearly identical (23.0 vs. 22.8 tokens/s). This phase-dependent gap suggests that the two frameworks utilize NPU compute resources differently as the workload shape changes.

Our source-code inspection reveals the underlying hardware-utilization gap. The MatMul implementation in llama.cpp uses only the NPU's vector accelerator, HVX, leaving the matrix accelerator HMX unused. In contrast, the QNN MatMul used by GENIE maps large matrix multiplications to HMX. This distinction is substantial because our microbenchmarks in Section~\ref{sec:micro_bench} measure 7.96~TOPS of FP16 compute on HMX, compared with only 0.33~TOPS on HVX---a 24.1$\times$ difference in peak compute capability. Thus, llama.cpp's MatMul underutilizes the NPU by leaving its most powerful compute unit idle.

To quantify the resulting operator-level gap, we sweep the MatMul output width $N$, which corresponds to the token-parallel width in prefill and is fixed to one in decode. Figure~\ref{fig:matmul_cycles} shows a clear turning point: at decode-like sizes ($N \le 4$), llama.cpp is comparable to QNN. At $N=256$, however, QNN reduces MatMul cycles by $26.9\times$--$30.7\times$; at $N=1024$, the reduction reaches $28.9\times$--$48.0\times$.

Prefill exposes large-$N$ GEMM; QNN scales efficiently in this regime, and GENIE inherits that advantage. Decode exposes small-$N$ GEMV, where the HMX has no advantage, so llama.cpp remains competitive. Thus, how to effectively utilize the NPU's matrix accelerator is a key factor in framework gaps, especially in prefill.

\begin{figure}[t]
  \centering
  \begin{subfigure}[t]{0.45\linewidth}
    \centering
    \includegraphics[width=\linewidth]{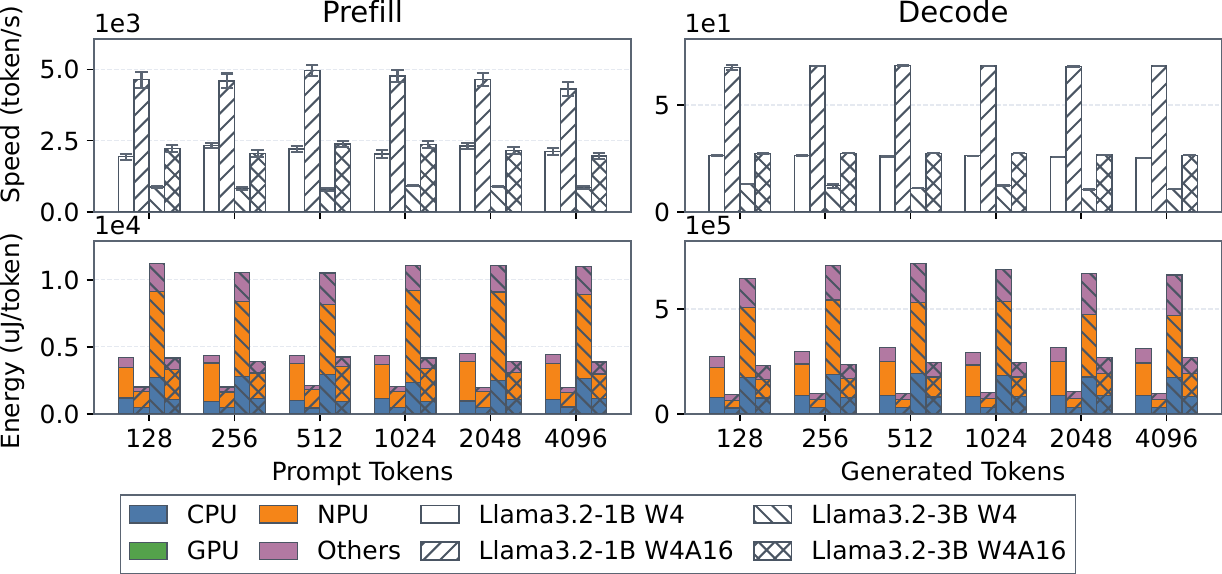}
    \caption{End-to-end impact.}
    \label{fig:activation_quantization}
  \end{subfigure}
  \hfill
  \begin{subfigure}[t]{0.53\linewidth}
    \centering
    \includegraphics[width=\linewidth]{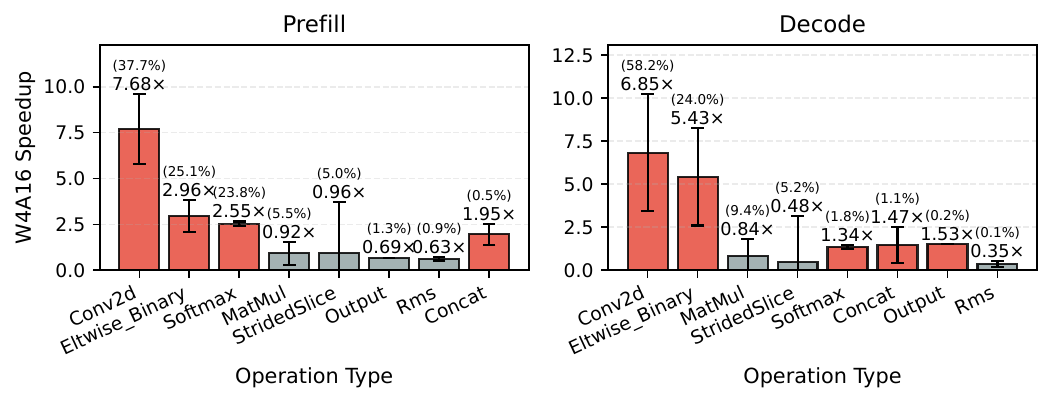}
    \caption{Operator breakdown.}
    \label{fig:activation_quant_operator_breakdown_3b}
  \end{subfigure}
  \caption{Impact of activation quantization in GENIE: (a) throughput and energy improvements from W4A16 over W4; (b) operator speedup breakdown for W4 vs. W4A16 on Llama-3.2-3B, where percentages denote each operator's contribution to total compute cycles.}
\end{figure}

\begin{table}[t]
\centering
\caption{Best perplexity (lower is better) across CPU, GPU, and NPU backends on WikiText-2. Values in parentheses report the PPL increase relative to the best CPU-W4 result for the same model.}
\label{tab:perplexity_w4_vs_w4a16}
\scriptsize
\setlength{\tabcolsep}{10pt}
\begin{tabular}{lcccc}
\toprule
Model & CPU-W4 & GPU-W4 & \textbf{NPU-W4} & \textbf{NPU-W4A16} \\
\midrule
Qwen2.5-1.5B & {18.11 (+0.0\%)} & {18.01 (-0.5\%)} & {18.17 (+0.4\%)} & {22.96 (+26.8\%)} \\
Qwen2.5-7B & {13.59 (+0.0\%)} & {13.87 (+2.1\%)} & {$>1000$} & {16.27 (+19.7\%)} \\
Llama3.2-1B & {26.75 (+0.0\%)} & {26.81 (+0.2\%)} & {26.79 (+0.2\%)} & {37.89 (+41.6\%)} \\
Llama3.2-3B & {20.15 (+0.0\%)} & {20.18 (+0.2\%)} & {20.15 (+0.0\%)} & {23.62 (+17.2\%)} \\
Phi-3.5-mini & {11.42 (+0.0\%)} & {11.42 (+0.0\%)} & {$>1000$} & {12.57 (+10.1\%)} \\
\bottomrule
\end{tabular}
\end{table}

\subsection{Finding 3: Activation Quantization is Powerful but Risky}
Activation quantization is the third major source of framework divergence. Unlike weight-only W4 quantization, whose accuracy is broadly comparable across CPU and NPU execution, activation quantization introduces a different tradeoff: it provides large speedups on NPUs, but can also create substantial accuracy risk.
For example, GENIE's W4A16 path uses a static per-tensor activation quantization scheme, which is an accuracy-unfriendly point in the activation-quantization design space. During model conversion, a calibration dataset is used to estimate activation ranges, and the resulting affine encodings are stored as fixed scale and offset values. At runtime, the backend reuses these precomputed encodings rather than adapting them to the current input. Moreover, each activation tensor shares one scale and one offset; for example, a hidden-state activation with shape $128\times 2048$ would quantize 262,144 values to INT16 using a single encoding. This coarse granularity is especially sensitive to outliers, since one extreme value can determine the quantization range for the entire tensor.

\textbf{Speedup analysis:} On GENIE, comparing W4 (4-bit weight quantization) with W4A16 (4-bit weight and 16-bit activation quantization) for Llama-3.2-1B and Llama-3.2-3B shows consistent end-to-end gains. As shown in Figure~\ref{fig:activation_quantization}, W4A16 improves throughput by $2.16\times$ and $2.58\times$ in prefill and by $2.63\times$ and $2.34\times$ in decode, while reducing energy per token by $53.5\%$--$66.9\%$.

The speedup is concentrated rather than uniform. Figure~\ref{fig:activation_quant_operator_breakdown_3b} shows that in Llama-3.2-3B, the gains are dominated by a few hot operators: in prefill, Conv2d, Eltwise\_Binary, and Softmax account for 86.6\% of W4 cycles and all accelerate substantially; in decode, Conv2d and Eltwise\_Binary alone account for 82.2\% of cycles and dominate the gain.

\textbf{Accuracy loss analysis:} 
Perplexity, where lower is better, reflects the model's language modeling quality. Table~\ref{tab:perplexity_w4_vs_w4a16} shows that the best W4 accuracy on CPU, GPU, and NPU is nearly identical for supported models. {The $>1000$ NPU-W4 entries reflect numerical failures in llama.cpp's Hexagon execution path, while the same quantized models give consistent CPU/GPU outputs. For Phi-3.5-mini and Qwen2.5-7B, we identified an out-of-bounds DMA read in the process.} {The W4A16 configurations have 10.1\%--41.6\% higher PPL than the CPU-W4 baselines. Within GENIE, W4A16 increases PPL by 20.7\% on Llama-3.2-1B and 4.7\% on Llama-3.2-3B relative to its NPU-W4 configurations.} Thus, activation quantization can remove a substantial performance bottleneck on NPUs, but only when model accuracy is robust enough to tolerate it.

\subsection{Summary}

RQ1 shows that the three key causes of framework gaps are partial offloading, which amplifies communication and memory overhead; hardware underutilization in large operators; and differences in activation-quantization support. Together, these mechanisms explain why frameworks on the same hardware with the same kernels can still differ by about $2\times$, and why custom-kernel paths can fall over $10\times$ behind vendor-optimized stacks in prefill.
Among these factors, activation quantization unlocks an NPU-specific acceleration path by leveraging dedicated INT8 accelerators, providing an additional $2\times$ throughput improvement. However, the resulting accuracy degradation remains an open challenge.

\section{RQ2: Backend Optimization}
\label{sec:rq2}

\begin{figure}[t]
  \centering
  \includegraphics[width=1\linewidth]{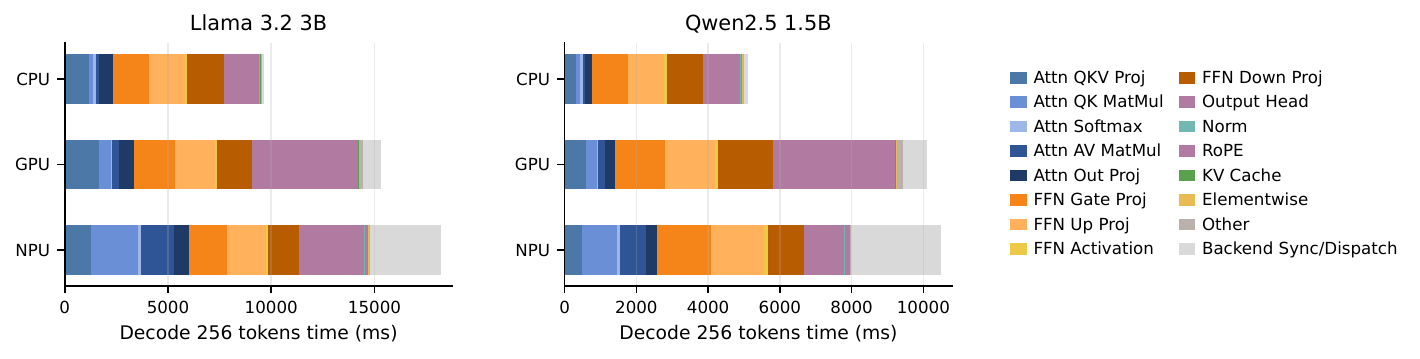}
  \captionsetup{skip=-2pt}
  \caption{CPU, GPU, and NPU decode-time breakdown.}
  
  \label{fig:time_breakdown}
\end{figure}

CPU, GPU, and NPU backends expose different execution models, leading to distinct tradeoffs across inference stages. We compare them to identify backend-specific strengths and bottlenecks.
{To reduce performance differences caused by suboptimal framework implementations, we use an efficient available implementation for each backend whenever possible.}

\subsection{Finding 4: NPU Decode Weakness}
Although NPUs dominate compute-bound prefill~\cite{xuFastOndeviceLLM2024,haoScalingLLMTestTime2025}, this advantage disappears in decode.
We decompose decode time by logical module in Figure~\ref{fig:time_breakdown}. The figure shows that the NPU is particularly slow in the \texttt{MUL\_MAT}-based attention and output-head modules, whereas the CPU, GPU, and NPU are comparable in other modules.
To explain these differences, we further break down the operators in Table~\ref{tab:mulmat_role}. For the NPU, we additionally report two hardware metrics: instruction cycles per packet (cyc/pkt), where $1.0$ indicates full pipelining and larger values indicate stalls, and the CPU dispatch share of each operator. Several observations stand out.

\begin{table*}[t]
\centering
\caption{Execution times of \texttt{MUL\_MAT}-based operators during the decoding of 256 tokens. Each \texttt{MUL\_MAT} is a GEMV: a $d_\text{in}\times d_\text{out}$ operand matrix times a length-$d_\text{in}$ activation vector yields $d_\text{out}$ outputs. cyc/pkt denotes execution cycles per instruction packet, and disp\% denotes the CPU dispatch share.}
\label{tab:mulmat_role}
\scriptsize
\setlength{\tabcolsep}{4pt}
\begin{tabular}{l l rrr rr @{\hspace{1.5em}} l rrr rr}
\toprule
 & \multicolumn{6}{c}{\textbf{Llama-3.2-3B}} & \multicolumn{6}{c}{\textbf{Qwen2.5-1.5B}} \\
\cmidrule(lr){2-7}\cmidrule(lr){8-13}
Operator & $d_\text{in}{\times}d_\text{out}$ & \makecell{CPU\\(ms)} & \makecell{GPU\\(ms)} & \makecell{NPU\\(ms)} & cyc/pkt & disp\% & $d_\text{in}{\times}d_\text{out}$ & \makecell{CPU\\(ms)} & \makecell{GPU\\(ms)} & \makecell{NPU\\(ms)} & cyc/pkt & disp\% \\
\midrule
QKV Proj      & $3072{\times}3072$   & 1186 & 1642 & 1282 & 3.12 & \underline{\textbf{28}} & $1536{\times}1536$   &  329 &  582 &  489 & 3.30 & \underline{\textbf{51}} \\
QK MatMul     & $128{\times}256$     &  179 &  606 & 2262 & \underline{\textbf{8.99}} & \underline{\textbf{21}} & $128{\times}256$     &   98 &  317 &  977 & \underline{\textbf{7.43}} & \underline{\textbf{22}} \\
AV MatMul     & $256{\times}128$     &  143 &  318 & 1569 & \underline{\textbf{8.90}} & \underline{\textbf{28}} & $256{\times}128$     &   61 &  173 &  720 & \underline{\textbf{7.41}} & \underline{\textbf{29}} \\
Attn Out Proj & $3072{\times}3072$   &  678 &  744 &  732 & 3.02 & \underline{\textbf{19}} & $1536{\times}1536$   &  195 &  294 &  291 & 2.86 & \underline{\textbf{35}} \\
FFN           & $3072{\times}8192$   & 5333 & 5632 & 5244 & 3.03 &  9 & $1536{\times}8960$   & 3000 & 4319 & 4008 & 2.71 & 12 \\
\bottomrule
\end{tabular}
\end{table*}

\textbf{Small attention kernels expose NPU memory and pipeline inefficiencies.} The large QKV and attention output projections remain relatively close to CPU performance, whereas the small QK and AV matmuls account for the main NPU gap. Across the two models, QK is $9.9\times$--$12.7\times$ slower on the NPU than on the CPU, and AV is $10.9\times$--$11.7\times$ slower. These kernels have small shapes ($128{\times}256$ or $256{\times}128$) and repeatedly read the KV cache. Due to the limited memory read speed measured in Section~\ref{sec:micro_bench}, their parallelism is insufficient to hide the NPU's memory-read latency, resulting in frequent pipeline stalls: QK and AV reach 7.41--8.99 cyc/pkt, compared with only 2.71--3.30 cyc/pkt for the larger projection and FFN kernels. Thus, small kernel sizes and slow KV-cache reads cause the NPU's poor attention performance.

\textbf{The NPU pays a large dispatch tax.} Table~\ref{tab:mulmat_role} also shows that CPU-side dispatch is already a substantial fraction of individual NPU operators: it accounts for 9--28\% of operator time on Llama-3.2-3B and 12--51\% on Qwen2.5-1.5B, reaching 51\% for Qwen's QKV projection. Decode invokes about 533 operators per token, so the fixed cost accumulates rapidly. In particular, many element-wise operators, such as RMS\_NORM, ADD, and MUL, perform only 1--3\,\textmu s of useful computation, whereas each host--NPU round trip takes 20--30\,\textmu s. Consistent with this per-operator evidence, Figure~\ref{fig:time_breakdown} attributes 3449\,ms on the NPU, compared with 71\,ms on the CPU, to the aggregate Backend Sync/Dispatch category. Together, the measurements show that the dispatch overhead is significant.

\textbf{Static graph overhead.} Decode emits one token per step, exposing little parallelism while repeatedly reading the growing KV cache. The workload is therefore memory-bound rather than compute-bound. In this regime, the NPU's peak compute capability is underutilized. The NPU further amplifies the problem by executing decode with shape-specialized static graphs. To support longer decoding, NPU frameworks typically set a large context size (e.g., 4096), making each step more expensive.
\begin{figure}[t]
  \centering
  \includegraphics[width=0.6\linewidth]{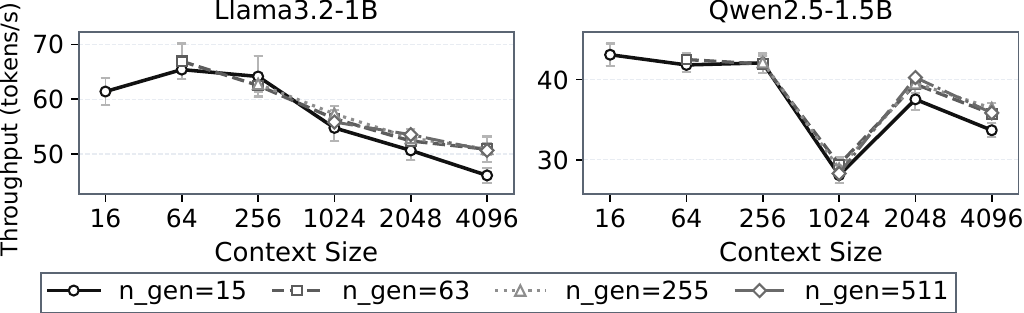}
  \caption{Decode throughput of the QNN NPU backend across context sizes. Throughput mainly depends on context size.}
  \label{fig:mllm_decode_ctx_throughput}
\end{figure}
Figure~\ref{fig:mllm_decode_ctx_throughput} shows that NPU decode is governed mainly by context size rather than output length. Throughput drops from 61.6 to 47.0 tokens/s for Llama-3.2-1B and from 43.9 to 35.1 tokens/s for Qwen2.5-1.5B as context grows from 16 to 4096 tokens, i.e., by 23.8\% and 20.0\%, respectively. By contrast, at fixed context size, the curves for 15, 63, and 255 generated tokens nearly overlap, showing that the per-token cost is set by context length rather than the remaining decode length. However, the NPU's static-graph design requires preallocating a large context size (e.g., 4096) to support future tokens, so each decode step incurs the cost of this large context even when the current sequence is short. This explains why context size, not the number of generated tokens, dominates NPU decode throughput.

\begin{figure*}[t]
  \centering
  \includegraphics[width=\linewidth]{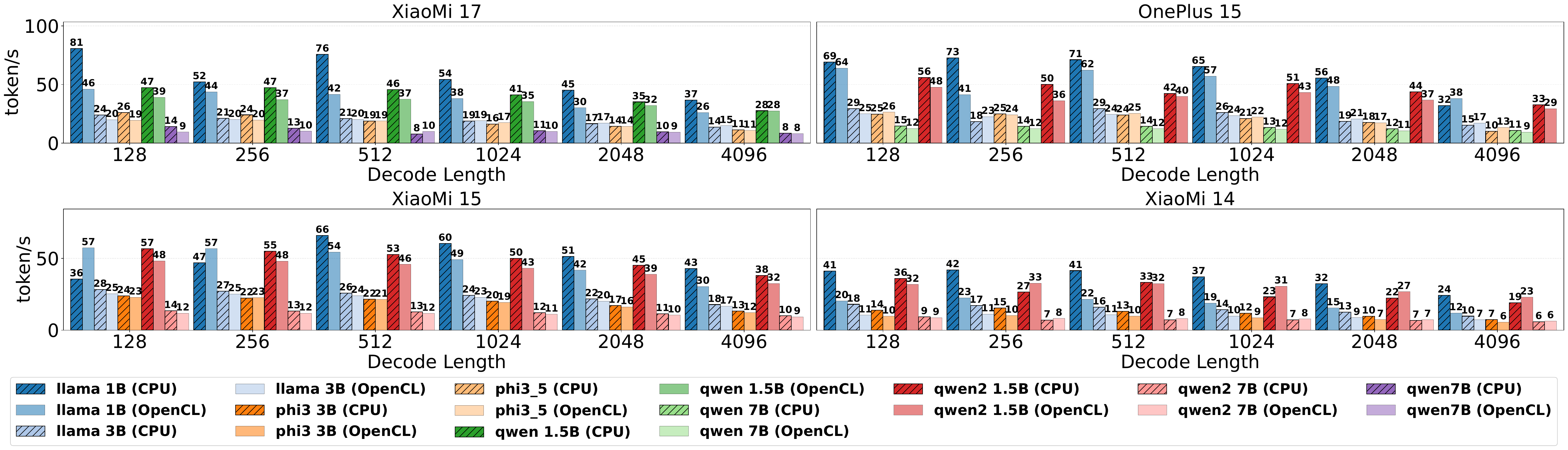}
  \caption{CPU and GPU decode throughput across devices for each model.}
  \label{fig:decode_speed_4devices}
\end{figure*}

\subsection{Finding 5: CPU Backends Dominate Decode Throughput}
Across the four devices, CPU backends generally deliver the highest decode throughput. Figure~\ref{fig:decode_speed_4devices} shows that CPU usually outperforms GPU across model families and configurations; GPU wins only in a few isolated cases, which do not overturn the overall trend.

More importantly, the GPU is often a near-equivalent substitute for decode throughput. As decode length grows, the CPU advantage typically narrows, while sustained decoding on the CPU increases occupancy and tightens CPU-side scheduling constraints. Offloading decode to the GPU can therefore preserve near-CPU throughput while reducing CPU pressure, making the GPU a practical auxiliary backend when CPU resources are scarce or reserved for concurrent tasks. With appropriate scheduling, the GPU can also be more energy-efficient in some regimes; we analyze these cases in Sections~\ref{sec:thread-core} and~\ref{sec:dvfs}.

\subsection{Summary}
RQ2 reveals strong phase and length dependence. The NPU is best for compute-bound prefill, but decode shifts to an invocation- and bandwidth-dominated regime in which static-graph overhead and poor amortization expose the NPU's main bottleneck. The CPU is usually the best decode backend because of its low dispatch overhead and effective cache use, while the GPU becomes attractive for long-context decode when it can deliver near-CPU throughput with lower CPU pressure and sometimes better energy efficiency.

\section{RQ3: Scheduling Strategy}
\label{sec:rq3}
\begin{figure}[t]
  \centering
  \includegraphics[width=\linewidth]{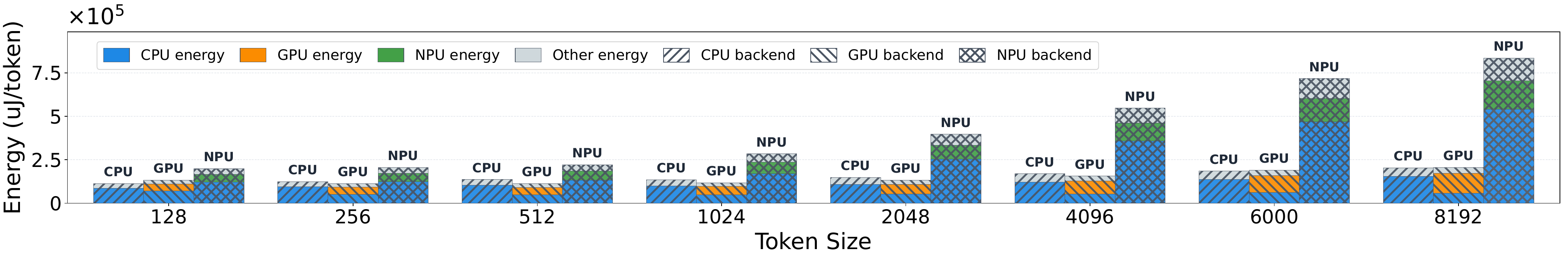}
  \captionsetup{skip=-2pt}
  \caption{Backend decode energy on Xiaomi 17 with Qwen2.5-1.5B.}
  
  \label{fig:decode_energy}
\end{figure}

Figure~\ref{fig:decode_energy} shows that even when computation is offloaded to the GPU or NPU, the host CPU still accounts for a substantial share of total system energy due to dispatch, polling, synchronization, and runtime coordination. We study this scheduling problem from two perspectives: inter-backend coordination between the CPU and accelerator, including RPC polling, NPU sleep control, and CPU frequency selection for GPU/NPU backends; and intra-backend tuning within each processor, including thread--core affinity and backend-local DVFS. Prior work~\cite{huangMNNAECSEnergyOptimization2025,zhangDissectingImpactMobile2025a} has focused mainly on CPU and GPU scheduling, leaving NPU-specific optimization and cross-unit interactions underexplored.

\begin{figure*}[t]
  \centering
  \includegraphics[width=\linewidth]{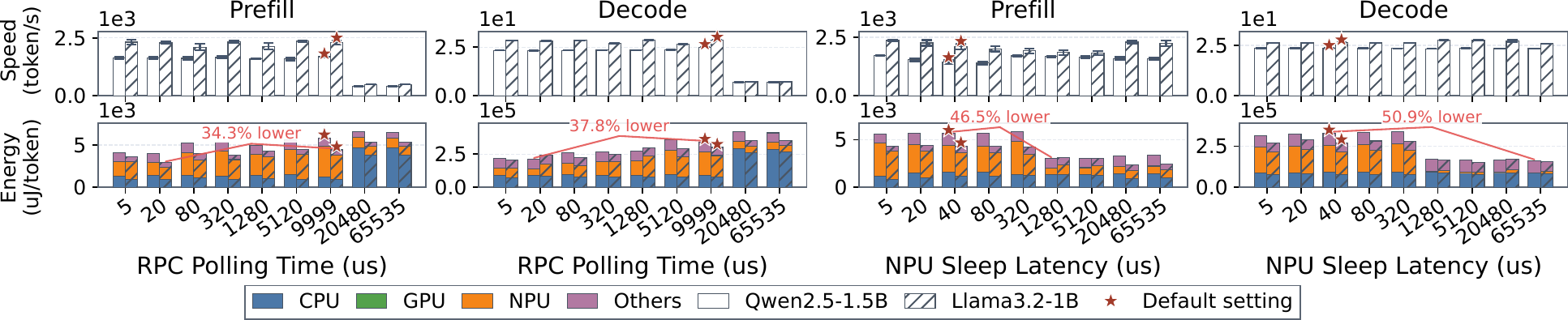}
  \caption{Impact of CPU--NPU invocation parameters on throughput and energy consumption.}
  \label{fig:npu_invocation}
\end{figure*}

\subsection{Finding 6: NPU Invocation Optimization}\label{sec:npu-invocation}

We first examine inter-backend coordination for the NPU backend. In modern mobile SoCs, the CPU and NPU operate independently and communicate through remote procedure calls (RPC). During LLM inference, the CPU sends parameters via RPC messages and the NPU executes operations; return values are sent back asynchronously. This design makes end-to-end efficiency sensitive to two host--accelerator control parameters: RPC polling interval and NPU sleep latency.

\textbf{RPC polling interval.} Figure~\ref{fig:npu_invocation} sweeps the RPC polling interval from 5 to 65535 $\mu$s, with 9999 $\mu$s as the default setting in GENIE and MLLM. Across both models and both phases, throughput is largely insensitive to this setting, and moving from the default to the energy-optimal setting changes throughput by only $-7.2\%$ to $+0.1\%$. The effect on energy is much stronger. Switching from the default to short polling intervals (5--20 $\mu$s) reduces total energy per token by $30.9\%$--$37.8\%$. The dominant source of this gain is the NPU rather than the host CPU. The NPU energy component drops by $49.8\%$--$70.8\%$, whereas the CPU component changes only modestly. This indicates that faster completion detection shortens the time the NPU remains unnecessarily active, while the extra host polling overhead is very small.

\textbf{NPU sleep latency.} Figure~\ref{fig:npu_invocation} also sweeps NPU sleep latency from 5 to 65535 $\mu$s, with GENIE and MLLM defaulting to 40 $\mu$s. Here, the default is too aggressive for energy efficiency. The energy-optimal points shift to much larger values, between 1280 and 65535 $\mu$s depending on model and phase, and can reduce total energy per token by $44.6\%$--$50.9\%$. Moreover, this saving does not sacrifice throughput, which ranges from a negligible $0.9\%$ drop to a $16.2\%$ increase; for Llama decode, the same setting is both throughput-optimal and energy-optimal. The reduction is driven primarily by the accelerator itself, whose NPU energy component falls by $66.7\%$--$100.0\%$. The overall pattern suggests that, for mobile LLM inference, allowing the NPU to remain idle longer before wake/sleep transitions yields a substantially better efficiency point.

\subsection{Finding 7: Thread--Core Affinity Matters} \label{sec:thread-core}
We find that CPU thread--core affinity is not only an important configuration for the CPU backend itself, but also a major factor shaping efficiency for GPU and NPU execution.
\begin{figure*}[t]
  \centering

  \begin{subfigure}[t]{0.32\textwidth}
    \centering
    \includegraphics[width=\linewidth]{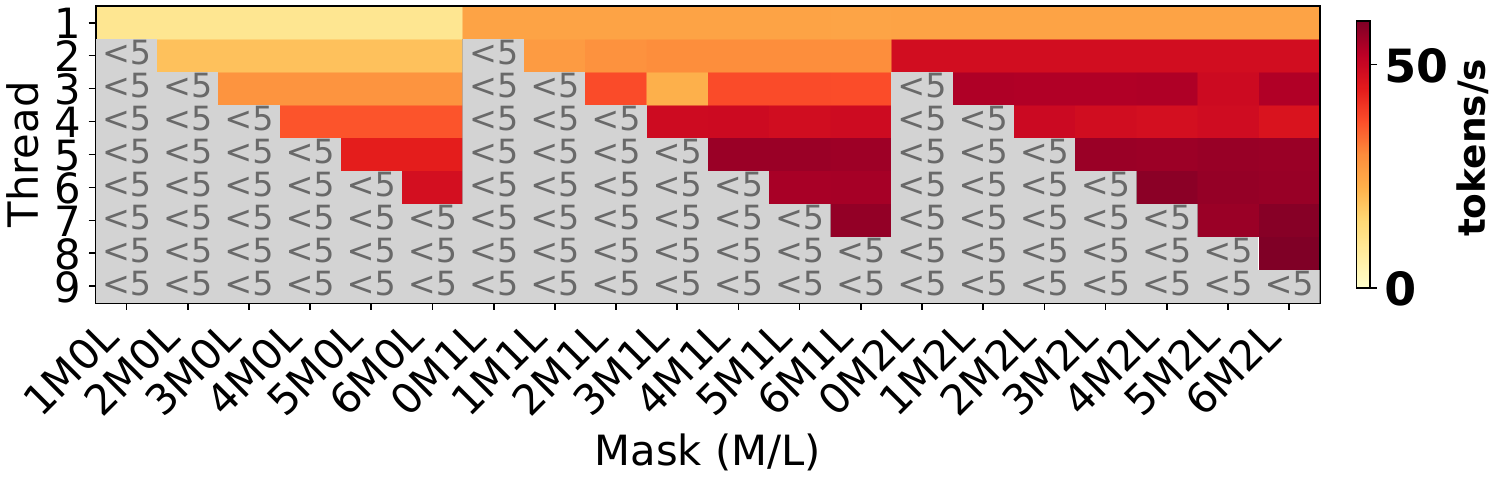}
    \caption{CPU: speed.}
    \label{fig:cpu_mask_speed_cpu}
  \end{subfigure}\hfill
  \begin{subfigure}[t]{0.32\textwidth}
    \centering
    \includegraphics[width=\linewidth]{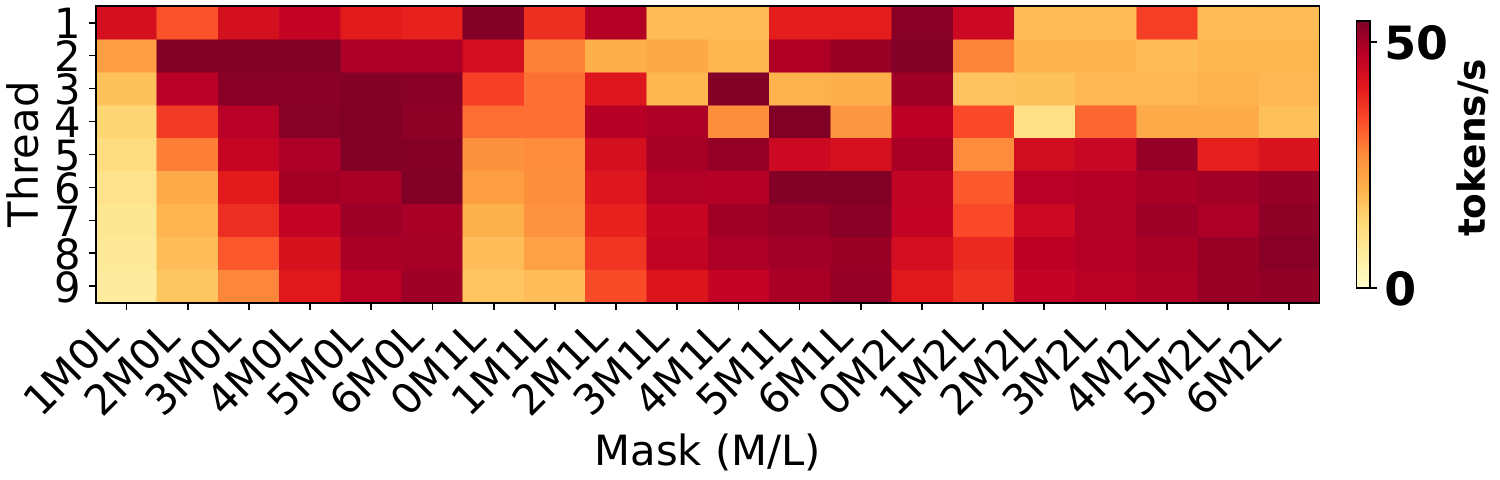}
    \caption{OpenCL: speed.}
    \label{fig:cpu_mask_speed_opencl}
  \end{subfigure}\hfill
  \begin{subfigure}[t]{0.32\textwidth}
    \centering
    \includegraphics[width=\linewidth]{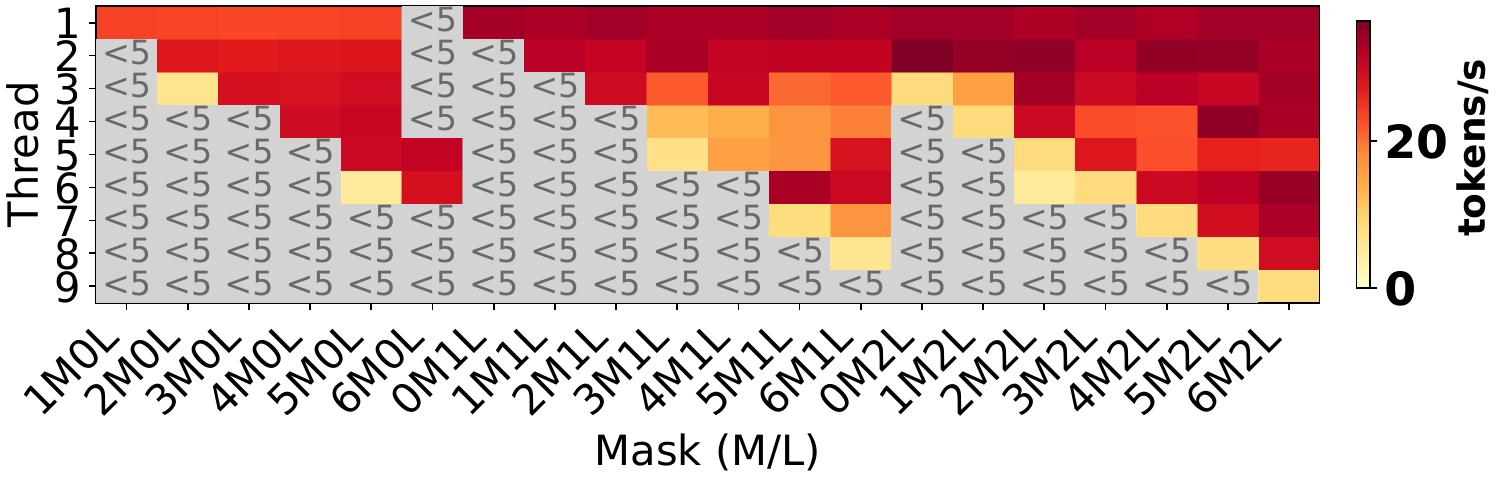}
    \caption{NPU: speed.}
    \label{fig:cpu_mask_speed_htp}
  \end{subfigure}

  \begin{subfigure}[t]{0.32\textwidth}
    \centering
    \includegraphics[width=\linewidth]{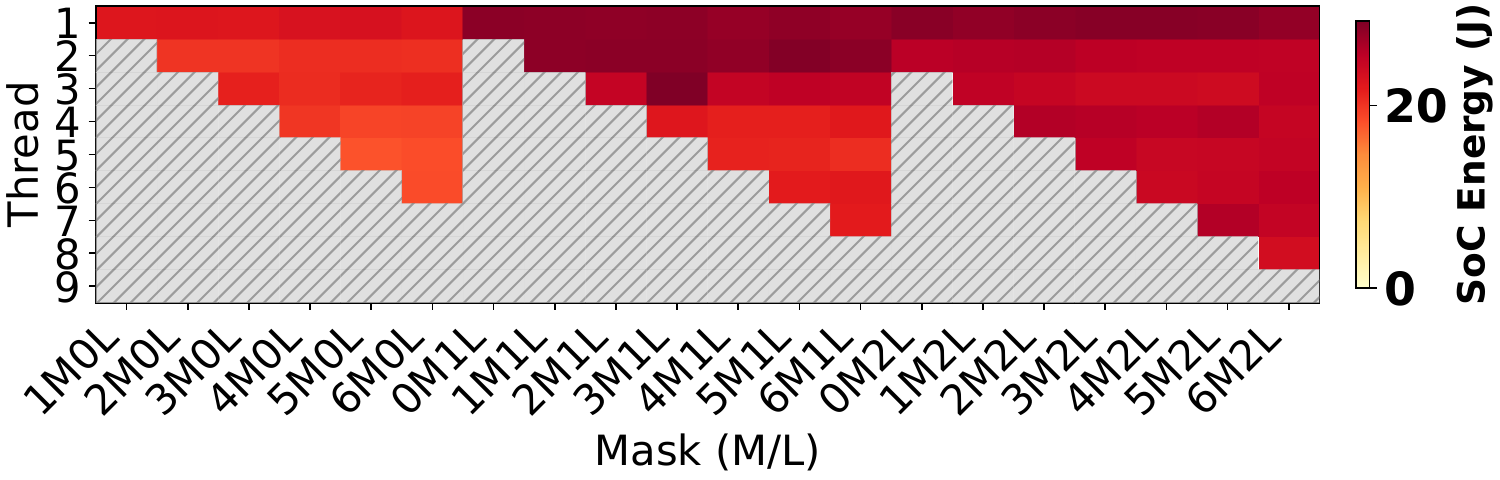}
    \caption{CPU: energy.}
    \label{fig:cpu_mask_energy_cpu}
  \end{subfigure}\hfill
  \begin{subfigure}[t]{0.32\textwidth}
    \centering
    \includegraphics[width=\linewidth]{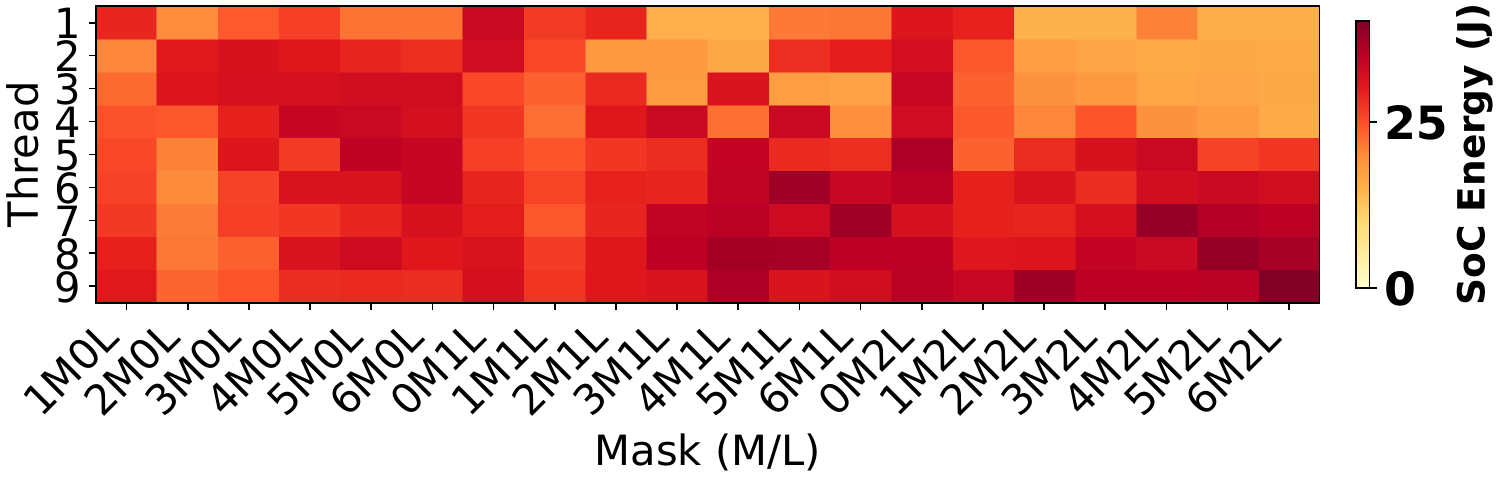}
    \caption{OpenCL: energy.}
    \label{fig:cpu_mask_energy_opencl}
  \end{subfigure}\hfill
  \begin{subfigure}[t]{0.32\textwidth}
    \centering
    \includegraphics[width=\linewidth]{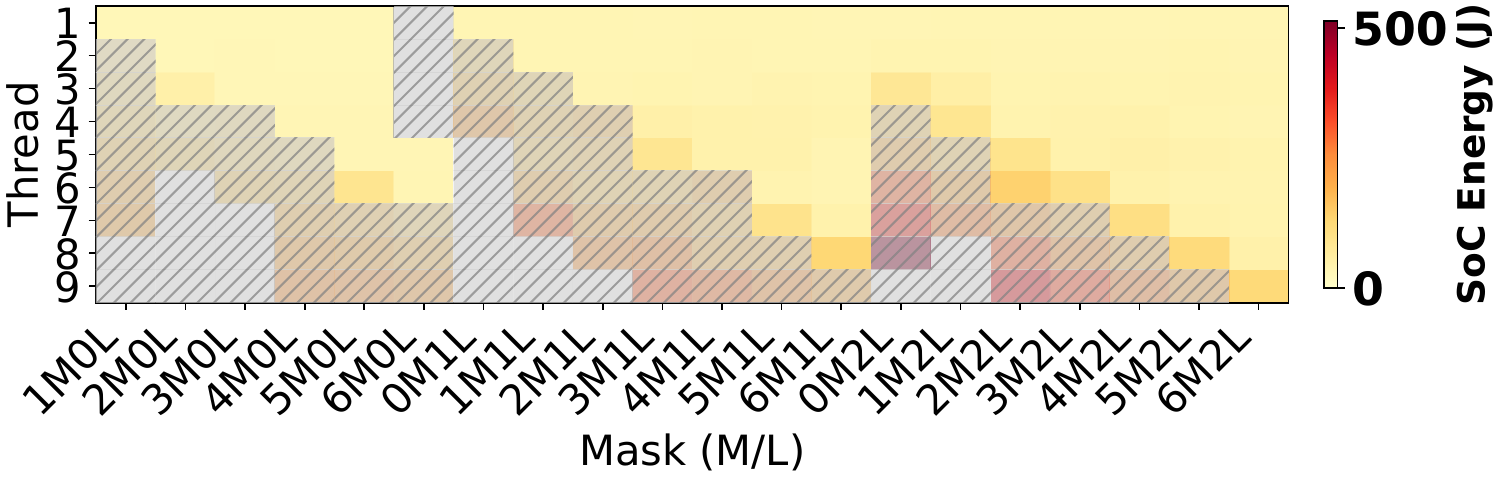}
    \caption{NPU: energy.}
    \label{fig:cpu_mask_energy_htp}
  \end{subfigure}

  \caption{Impact of CPU thread masking on throughput (top row) and system energy (bottom row) across backends for Qwen2.5-1.5B decode on Xiaomi 17.}
  \label{fig:cpu_mask_speed_energy_2x3}
\end{figure*}

\textbf{CPU and NPU backends favor matched thread and core counts.} For both the CPU and NPU backends, blindly increasing the thread count can significantly degrade decode throughput. As shown in Figures~\ref{fig:cpu_mask_speed_cpu} and~\ref{fig:cpu_mask_speed_htp}, performance is typically highest when the thread count roughly matches the number of active cores; once the thread count exceeds core availability, throughput drops sharply and can even become impractically low. In other words, what matters is not core selection alone, but thread--core co-design.

\textbf{NPU execution prefers minimal host participation.} Figures~\ref{fig:cpu_mask_speed_htp} and~\ref{fig:cpu_mask_energy_htp} further show that the NPU backend achieves its best throughput and energy efficiency with the smallest CPU thread count. Additional CPU threads mainly add orchestration overhead and contention, rather than useful parallelism.

\textbf{GPU execution needs enough, but not excessive, host concurrency.} The GPU backend exhibits a more nuanced pattern. As shown in Figures~\ref{fig:cpu_mask_speed_opencl} and~\ref{fig:cpu_mask_energy_opencl}, setting the CPU thread count above the number of active cores has limited impact on throughput and can even reduce energy in some cases, but using too few threads can significantly hurt GPU throughput. This suggests that GPU execution benefits from sufficient host-side concurrency for dispatch and synchronization, yet does not follow the same ``more threads is better'' logic as CPU-only execution.

Taken together, Figure~\ref{fig:cpu_mask_speed_energy_2x3} shows that thread affinity is a system-level scheduling knob rather than a CPU-only optimization. Even when the GPU or NPU performs most arithmetic work, the host-side thread configuration still shapes end-to-end throughput and energy.
\subsection{Finding 8: Energy-Inefficient DVFS}\label{sec:dvfs}

We next examine DVFS from both backend-local and cross-backend perspectives. We sweep CPU and GPU frequency levels, and we sweep NPU operating points over the joint NPU core/bus voltage-corner space. Across these settings, the defaults are often not energy-optimal. They can over-provision host compute for accelerator-heavy workloads, miss better mid-range operating points, and ignore phase-specific behavior.

\textbf{Available frequency scales.} CPU cluster policy 0 (cpu0--cpu5) and cluster policy 6 (cpu6--cpu7) expose different frequency ranges. Because of vendor-enforced DVFS behavior, some frequency points cannot be fixed or maintained. Accordingly, we use 21 valid CPU frequency-level combinations and 15 GPU frequency levels in our evaluation. The NPU frequency states are configured at the governor level and include \textit{disable}, \textit{min}, \textit{svs2}, \textit{svs}, \textit{svs\_plus}, \textit{nom}, \textit{nom\_plus}, \textit{turbo}, \textit{turbo\_plus}, \textit{turbo\_l2}, \textit{turbo\_l3}, and \textit{max}.

\begin{figure*}[t]
  \centering

  \begin{subfigure}[t]{0.32\textwidth}
    \centering
    \includegraphics[width=\linewidth]{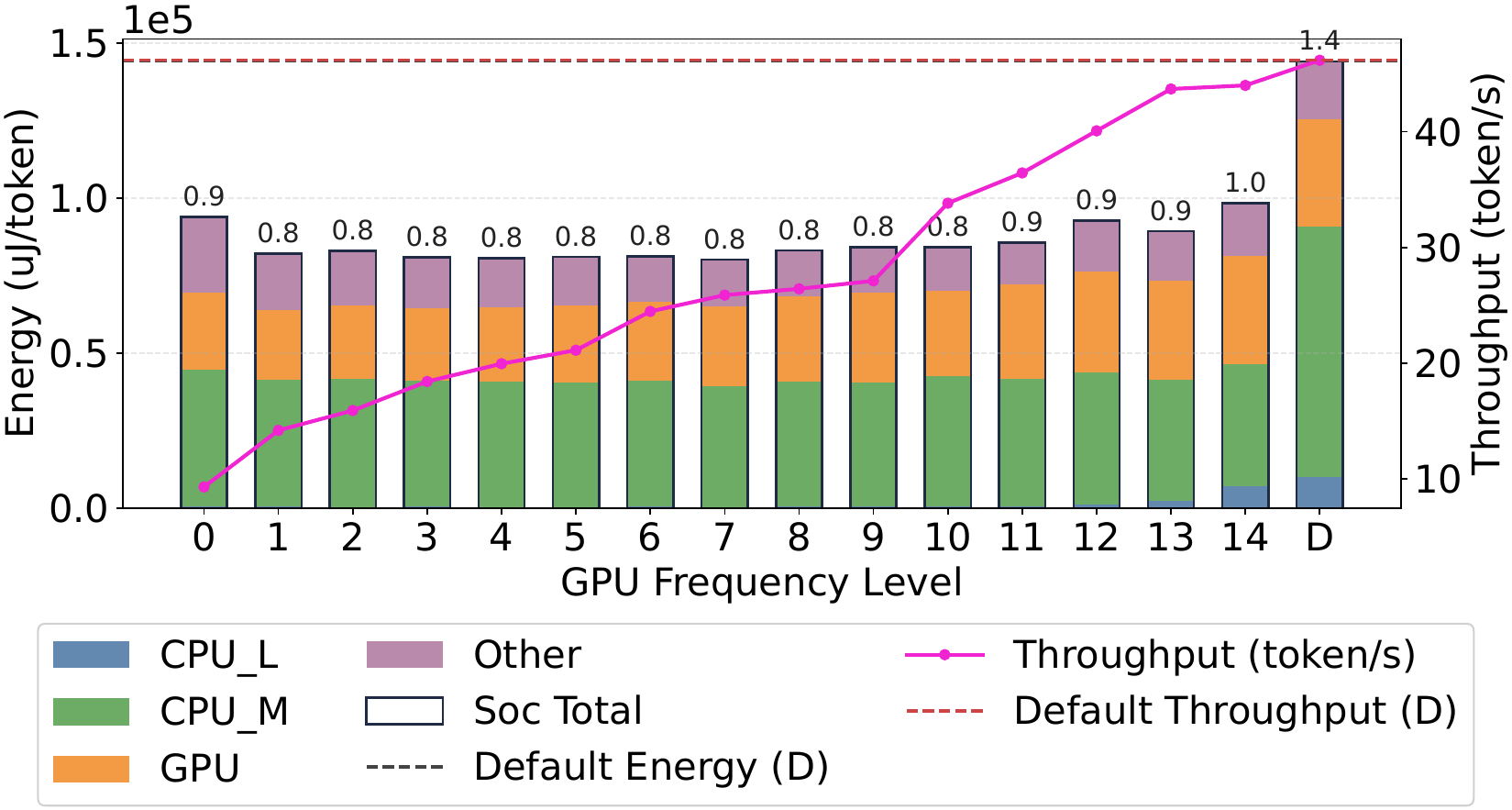}
    \caption{GPU frequency scaling.}
    \label{fig:GPU_freq_llama}
  \end{subfigure}
  \hfill
  \begin{subfigure}[t]{0.32\textwidth}
    \centering
    \includegraphics[width=\linewidth]{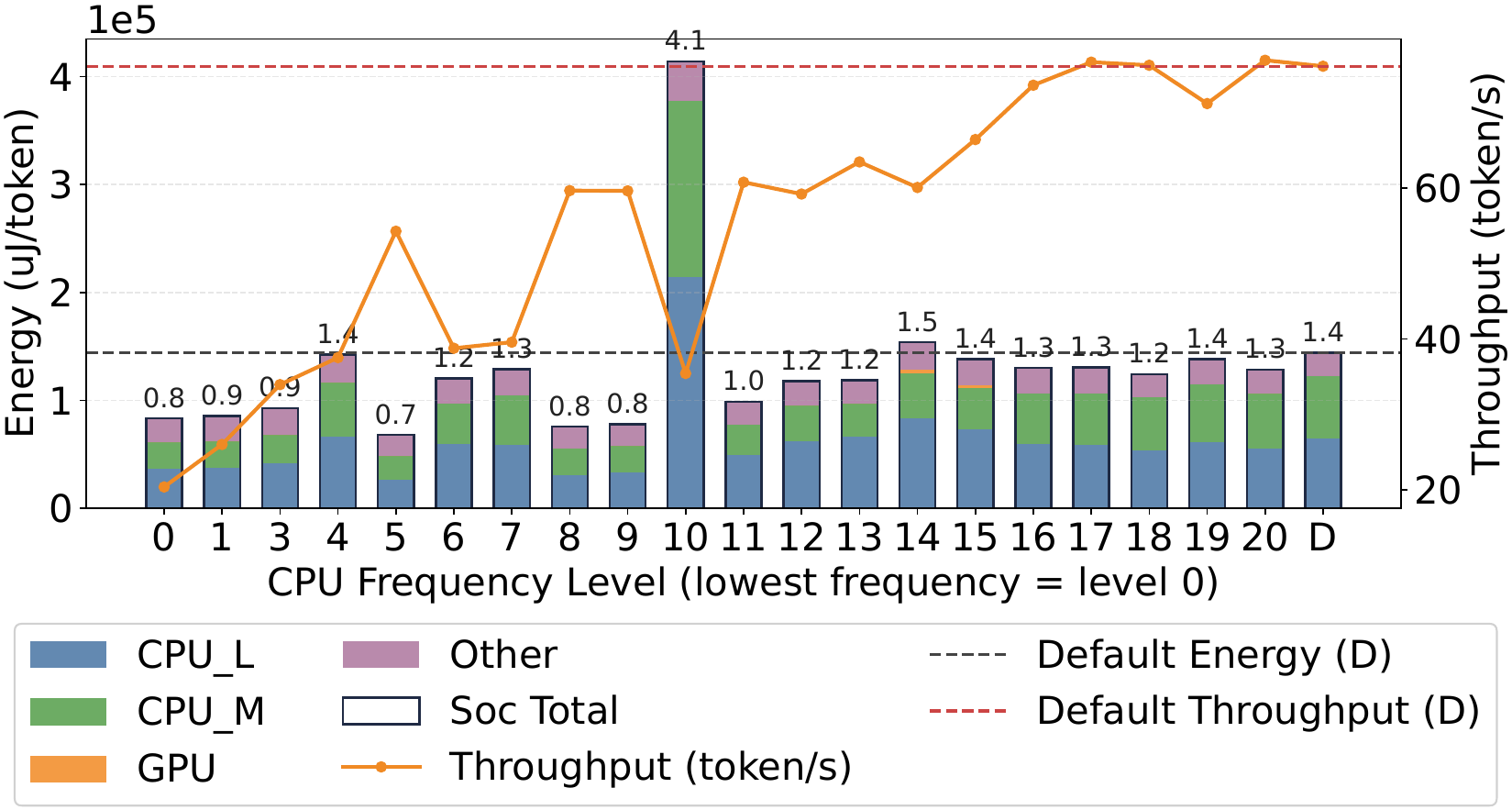}
    \caption{CPU frequency scaling.}
    \label{fig:CPU_freq_llama}
  \end{subfigure}
  \hfill
  \begin{subfigure}[t]{0.32\textwidth}
    \centering
    \includegraphics[width=\linewidth]{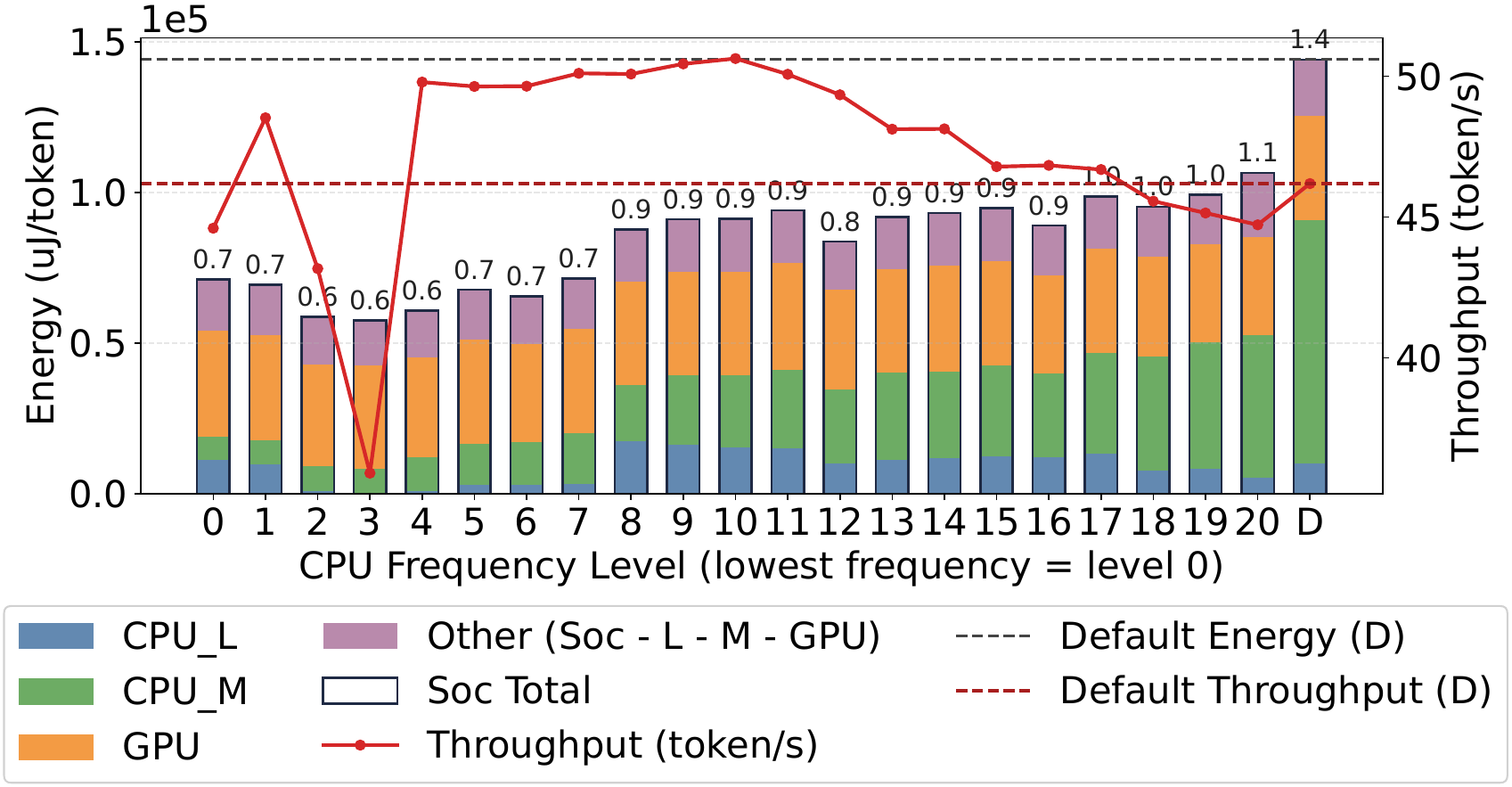}
    
   \caption{CPU frequency scaling for GPU.}
    \label{fig:cpu_freq_scaling_for_GPU}
  \end{subfigure}
  \captionsetup{skip=1pt} 
  \caption{Throughput and energy under frequency scaling in llama.cpp for Llama 3.2 1B decode with 256 tokens.}
  \label{fig:freq_scaling_combined}
\end{figure*}

\textbf{Backend-local DVFS.} Default backend-local DVFS policies leave substantial efficiency on the table, but the best response depends strongly on which processor is on the critical path.

For the CPU backend, Figure~\ref{fig:CPU_freq_llama} shows that the default DVFS setting is far from energy-optimal. Downclocking to frequency levels 8--9 reduces energy consumption by roughly 50\% relative to the default setting, with only about a 13\% drop in throughput. However, not all lower frequencies are beneficial: reducing the CPU frequency level from 20 to 14 incurs a 20\% throughput loss while increasing energy consumption by 14\%. In contrast, lowering the frequency level from 20 to 8 achieves a much better trade-off, reducing throughput by 20\% while cutting energy consumption by 42\%.

{For the GPU backend, moderate downclocking reduces energy consumption with a small throughput loss. Figure~\ref{fig:GPU_freq_llama} shows that level 13 reduces energy consumption by 38.0\% relative to the default setting, with a 5.4\% throughput drop. Further downclocking from level 13 to 10 reduces energy by another 5.9\% while lowering throughput by 22.6\%.}

\begin{figure*}[t]
  \centering
  \includegraphics[width=\linewidth]{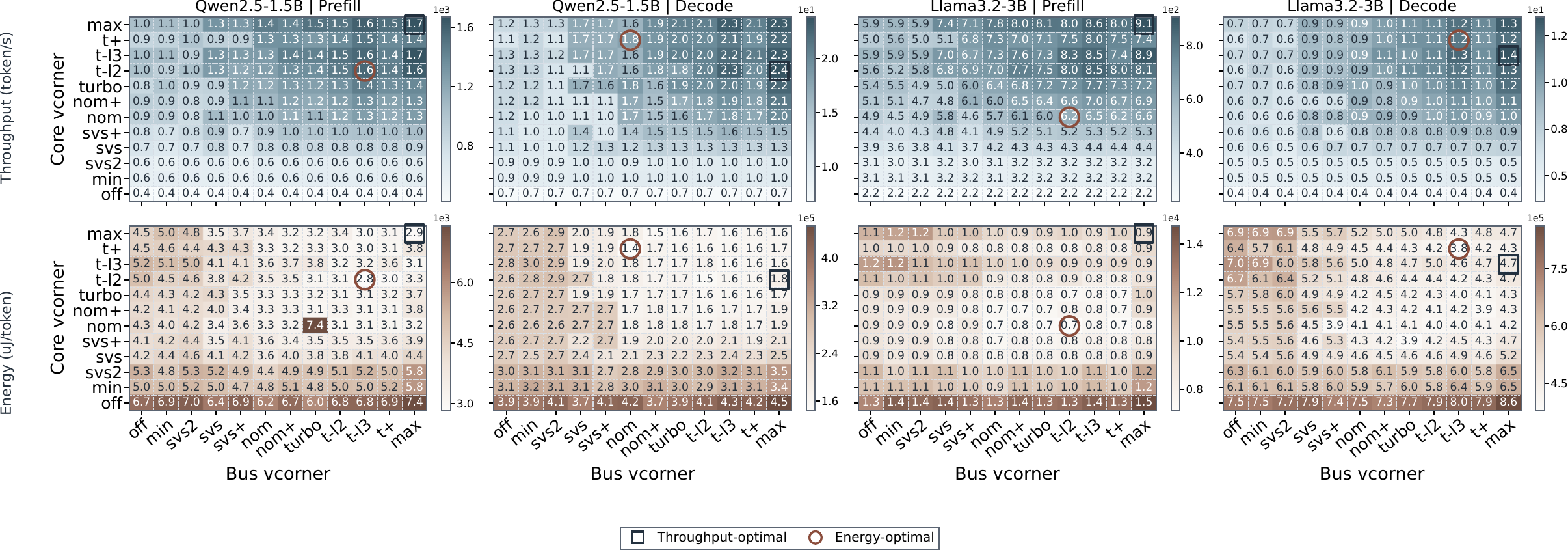}
  \caption{Impact of NPU voltage-corner scaling on throughput and energy during inference.}
  \label{fig:npu_vcorner_scaling_combined}
\end{figure*}

\begin{figure*}[t]
  \centering
  \includegraphics[width=\linewidth]{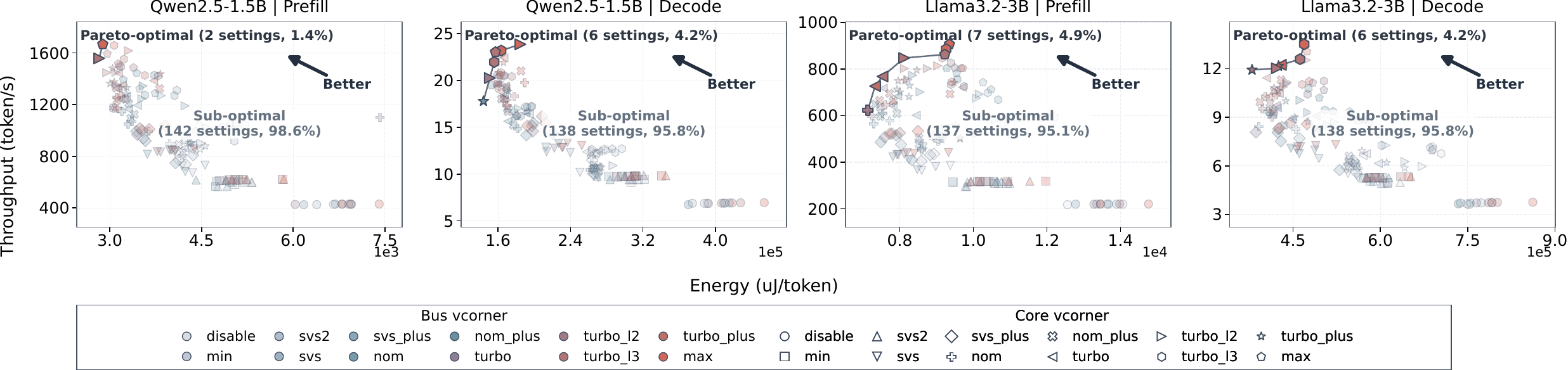}
  \caption{Throughput and energy trade-offs for NPU voltage-corner scaling. The Pareto-optimal settings are highlighted.}
  \label{fig:npu_vcorner_pareto_2x2}
\end{figure*}

NPU frequency scaling differs fundamentally from the CPU and GPU backends. While DVFS on CPU and GPU can be analyzed primarily along a single frequency axis, NPU scaling depends on the joint configuration of both NPU core and bus voltage corners. We therefore sweep the full core vcorner $\times$ bus vcorner space and visualize the results as a heatmap in Figure~\ref{fig:npu_vcorner_scaling_combined}; Figure~\ref{fig:npu_vcorner_pareto_2x2} further projects the same settings into the throughput--energy plane and highlights Pareto-optimal operating points. The results show that NPU DVFS is inherently a two-dimensional operating-point problem rather than a single monotonic knob, and they reveal three consistent patterns.

First, throughput-optimal points concentrate at very high voltage corners. This is expected because higher core and bus corners reduce compute and data-movement stalls, but the heatmap shows that this policy mostly optimizes speed rather than energy. In all four model--phase settings, the highest-throughput point lies near the upper-right region of the vcorner space, confirming that the NPU still benefits from aggressive provisioning when raw throughput is the only objective.

Second, energy-optimal points usually differ from throughput-optimal points, but the size of the trade-off depends strongly on model and phase. In Qwen prefill, the two optima are close: moving from (max, max) to the energy-optimal (turbo\_l2, turbo\_l3) lowers energy by only 2.0\% while reducing throughput by 6.5\%. In the other three cases, however, the separation is substantial. Qwen decode saves 22.1\% energy at the energy-optimal (turbo\_plus, nom) point, but throughput drops by 25.5\%. Llama prefill saves 23.7\% energy at (nom, turbo\_l2) with a 31.3\% throughput reduction, while Llama decode saves 19.1\% energy at (turbo\_plus, turbo\_l3) with an 11.7\% throughput drop. None of the energy-optimal points falls into min/svs/svs2; instead, they cluster around nom to turbo\_plus for the core and nom to turbo\_l3 for the bus. Thus, aggressive underclocking is not the right policy for these NPU workloads.

Third, most NPU vcorner settings are not useful operating points at all. In Figure~\ref{fig:npu_vcorner_pareto_2x2}, a Pareto-optimal setting means that no other measured setting can improve throughput without increasing energy, or reduce energy without lowering throughput. Only 2--7 out of 144 settings per panel are Pareto-optimal, and across all four panels 555 out of 576 settings (96.4\%) are suboptimal. The penalty is large: a dominated setting can be over 3.5$\times$ slower than another setting with no higher energy, or consume over 2$\times$ as much energy as another setting with no lower throughput; in throughput per unit energy, the gap reaches roughly 7--10$\times$. Therefore, NPU DVFS scheduling should select operating points from the corresponding Pareto frontier.

\textbf{Cross-backend DVFS.} Backend-local policies overlook strong CPU--accelerator interactions and can therefore miss globally efficient operating points.

CPU downclocking can benefit the GPU backend: Figure~\ref{fig:cpu_freq_scaling_for_GPU} shows that when running the GPU backend, a higher CPU frequency is not always better. At CPU frequency level 9, throughput increases by 13\% relative to the default setting, while energy consumption decreases by 51\%. More generally, mid-range CPU frequency levels (5--10) can outperform higher levels (20--26) in both throughput and energy, indicating that excessive host frequency can add overhead rather than useful work.

\begin{figure}[t]
  \centering
  \includegraphics[width=0.85\linewidth]{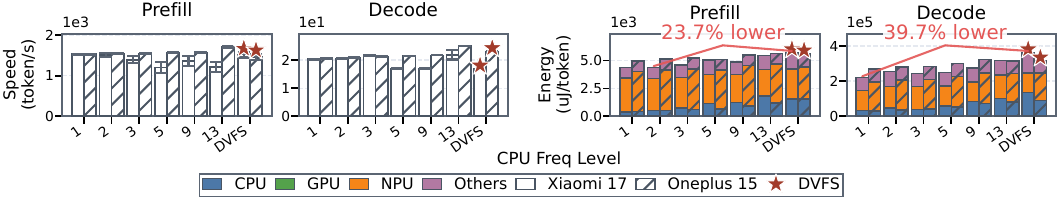}
  \caption{Impact of CPU frequency scaling on throughput and energy during NPU inference for Qwen2.5-1.5B, across prefill and decode on Xiaomi 17 and OnePlus 15.}
  \label{fig:cpu_freq_scaling}
\end{figure}

CPU frequency scaling strongly affects NPU efficiency: Figure~\ref{fig:cpu_freq_scaling} fixes the CPU at several frequency levels during NPU inference and compares them with the default DVFS setting. Across both \textit{prefill} and \textit{decode}, the results show that the default DVFS policy is not energy-optimal for NPU execution. Lower CPU frequency levels generally reduce total energy per token while preserving comparable throughput, indicating that host-side CPU provisioning under DVFS is often excessive for accelerator-dominated inference. This effect is especially clear in \textit{prefill}, where throughput remains largely unchanged while energy is consistently reduced, with the largest reduction reaching 23.7\%. A similar pattern also appears in \textit{decode}, where the best case achieves a 39.7\% energy reduction, although throughput varies more across frequency levels, suggesting stronger device- and configuration-dependent trade-offs.

Overall, these results demonstrate that DVFS substantially affects NPU and GPU inference. Unlike pure CPU execution, where lowering frequency often trades speed for energy, NPU- and GPU-bound workloads frequently tolerate substantial CPU downclocking because the critical path remains on the accelerator.

\subsection{Summary}
RQ3 shows that scheduling is critical to mobile LLM efficiency. The main gains come from reducing host-side overhead through better CPU--NPU invocation settings, proper thread--core affinity, and backend-aware DVFS. Overall, the best policy is phase- and backend-aware rather than relying on default settings.

\section{Optimization and Future Directions}
\label{sec:future}

Guided by these findings, we identify a practical energy-oriented configuration for NPU-based mobile LLM inference: use QNN with full-graph offloading to the NPU, build a computation graph with a context length appropriate for the target workload, and reduce both CPU- and NPU-side overhead by tuning the CPU polling interval, CPU frequency, and NPU sleep latency. 
{We estimate latency by dividing each phase's token count by its measured throughput, and SoC energy by multiplying it by the measured energy per token, then summing across prefill and decode.}

{Specifically, we set the RPC polling interval to 20\textmu s, the NPU sleep latency to 65535\textmu s, and fix the CPU frequency at the lowest level to avoid unnecessary CPU boosting during NPU execution. This cuts prefill energy by 52.9\% with a slight throughput gain of 0.8\%, and cuts decode energy by 53.2\% with a 13.6\% throughput loss. Table~\ref{tab:opt_compare_e2e} reports the estimated impact on the MathQA~\cite{amini-etal-2019-mathqa}, RoleBench~\cite{wang2024rolellm}, and LongBench v2~\cite{bai2024longbench2} datasets. Across these datasets, SoC energy decreases by 52.9\%--53.2\%. Latency rises by 15.7\% on MathQA and 15.5\% on RoleBench, but decreases by 0.8\% on the prefill-heavy LongBench v2 workload.}

\begin{table}[t]
\centering
\scriptsize
\caption{Estimated end-to-end impact of the energy-optimized GENIE configuration on the MathQA, RoleBench, and LongBench datasets.}
\label{tab:opt_compare_e2e}
\setlength{\tabcolsep}{10pt}
\begin{tabular}{@{}lrr@{}}
\toprule
Dataset & Latency & Energy \\
\midrule
MathQA & {174187.8 $\rightarrow$ 201547.6 s (+15.7\%)} & {1268786.1 $\rightarrow$ 594278.9 J (-53.2\%)} \\
RoleBench & {407484.9 $\rightarrow$ 470686.1 s (+15.5\%)} & {2972770.6 $\rightarrow$ 1392517.5 J (-53.2\%)} \\
{LongBench v2} & {89583.0 $\rightarrow$ 88902.7 s (-0.8\%)} & {737890.3 $\rightarrow$ 347817.7 J (-52.9\%)} \\
\bottomrule
\end{tabular}
\end{table}

Our findings reveal several optimization directions:

\textbf{Accurate activation quantization on NPUs.} Finding 3 shows that activation quantization can improve throughput and energy efficiency by 2$\times$, but accuracy loss remains an unresolved problem on NPUs. Current activation quantization uses a coarse-grained per-tensor approach. Finer-grained methods, such as per-channel or per-group quantization, may improve accuracy while preserving the speedup, although achieving this balance remains challenging. Another promising direction is to develop a more robust calibration dataset and method for estimating activation ranges, which can reduce the impact of outliers and improve the overall accuracy of activation quantization.

\textbf{Multi-graph switching on NPUs.} Finding 4 shows that NPU-based decoding suffers from static graph execution: frameworks often compile a large graph for long contexts even when the current sequence is short, creating unnecessary overhead. A promising direction is to pre-compile multiple graphs for different context windows and switch among them during decoding. The key challenge is efficient switching, including KV-cache reuse and state migration.

\textbf{Heterogeneous phase-aware pipelines.} Current frameworks usually bind inference to a single backend, while our results show that prefill and decode favor different processors. A promising direction is a runtime that dispatches prefill to the NPU and decode to the CPU. The key challenge is minimizing tensor copy and manipulation overhead.

\textbf{Phase- and model-aware NPU frequency scaling.} Our DVFS results suggest that NPU operating points should not be fixed globally. The best operating point depends on both phase and model. Future work may jointly tune NPU frequency with workload features to balance throughput and energy.

\textbf{Holistic resource scheduling.} CPU thread--core affinity, polling behavior, sleep settings, and DVFS materially affect NPU and GPU inference even when the CPU is not the main compute engine. This suggests that future systems should jointly coordinate resources rather than tuning only the active processor.

Overall, these findings suggest that mobile LLM inference still has substantial cross-layer optimization headroom spanning frameworks, backends, and scheduling strategies.

{\section{Assumptions and Limitations}
\label{sec:limitations}

\subsection{Applicability of Findings}

\textbf{Hardware Platforms.}
We expect the NPU prefill advantage to generalize to other mobile SoCs, including those from Apple and MediaTek, because their NPUs offer substantially greater matrix-compute capacity than their CPUs and GPUs. Decode performance is primarily determined by memory access, so backend rankings require measuring the effective DDR-access bandwidth of the CPU, GPU, and NPU on each SoC.

\textbf{Frameworks and Models.}
Operator offloading, hardware utilization, and quantization remain relevant across frameworks, including ExecuTorch~\cite{executorch2026}, and model families. Their performance effects can vary with operator implementations and model architectures. For example, attention designs that reduce KV-cache size or memory traffic can change decode bottlenecks and backend suitability.

\textbf{CPU Downclocking.}
CPU downclocking can save energy when accelerator (e.g., NPU) execution dominates inference, while the CPU frequency should still meet inference latency requirements and keep concurrent applications responsive.

\subsection{Performance with Concurrent Applications}

Concurrent applications can compete for compute resources and for memory bandwidth across backends. Sereno~\cite{xin2026sereno} shows that NPU memory-bandwidth contention can degrade foreground responsiveness even when LLM throughput changes little. Applications also reduce memory available for weights and KV caches. KV-cache growth with context length further reduces available memory and can limit deployable model size. These interactions can degrade performance under concurrent workloads.}

\section{Related Work}
\label{sec:related}

\textbf{On-device LLM measurement studies.} Prior work establishes key baselines for on-device LLM inference. MELTing Point~\cite{laskaridisMELTingPointMobile2024} and MobileAIBench~\cite{murthyMobileAIBenchBenchmarkingLLMs2024} benchmark mobile LLM performance and resource usage, while later studies analyze runtime bottlenecks, model scaling, SLM behavior, and multi-instance execution~\cite{wangLmMeterUnveilingRuntime2025,luDemystifyingSmallLanguage2025,li_large_2024,guo_large_2025}. These works show that mobile LLM inference is increasingly practical with suitable models and system settings. However, they usually cover only limited framework--backend combinations, focus mainly on CPU and GPU execution, and rely on coarse whole-device energy measurements, leaving NPU behavior, cross-framework differences, and backend-specific coordination overheads insufficiently understood.

\textbf{On-device LLM optimizations.} A large body of research improves LLM efficiency through model compression and runtime optimization, including quantization and compression~\cite{gptq,awq,sparsegpt}, efficient attention and KV-cache management~\cite{flashattention,flashattention2,ye2025flashinfer,vllmPagedAttention,h2o,ainslie-etal-2023-gqa_kvcache}, and speculative decoding~\cite{speculativeDecoding,xuEdgeLLMFastOnDevice2025}. On mobile platforms, recent systems further explore sparsity-aware execution, NPU offloading, heterogeneous GPU--NPU collaboration, memory-aware deployment, architecture co-design, and scheduling optimization~\cite{song2024powerinfer,xue2024powerinfer,EdgeMoE,xu2024wip,xuFastOndeviceLLM2024,chen_characterizing_2025,m2LLM,mobile_foundation_as_firmware,lpspec2025,yin2025dynamic,haoScalingLLMTestTime2025,huangMNNAECSEnergyOptimization2025,zhangDissectingImpactMobile2025a}. These studies show that substantial gains can come from optimizing specific mechanisms. However, they usually target one framework, one backend path, or one optimization dimension at a time. Our work instead provides a benchmark and a unified cross-layer measurement view across five frameworks, three backends, and systematic scheduling policies, enabling us to jointly analyze framework diversity, phase-dependent backend behavior, and fine-grained energy inefficiency.

\section{Conclusion}
\label{sec:conclusion}

In this paper, we presented the first comprehensive cross-layer measurement study of mobile LLM inference across five representative frameworks and three heterogeneous backends. We also developed \texttt{PowerBench}, a lightweight profiling library that enables backend-specific throughput and energy attribution beyond coarse device-level measurements. Our findings showed how framework design, backend optimization, and resource scheduling jointly determine end-to-end mobile LLM efficiency. More importantly, we revealed substantial opportunities for efficiency improvements, especially in NPU execution.

\begin{acks}
This work is in part supported by National Natural Science Foundation of China 62525208, U22A2031, 62402028, and Beijing Natural Science Foundation (L2603011).
\end{acks}

\bibliographystyle{unsrtnat}
\bibliography{refs}

@inproceedings{ye2025flashinfer,
  author       = {Zihao Ye and
                  Lequn Chen and
                  Ruihang Lai and
                  Wuwei Lin and
                  Yineng Zhang and
                  Stephanie Wang and
                  Tianqi Chen and
                  Baris Kasikci and
                  Vinod Grover and
                  Arvind Krishnamurthy and
                  Luis Ceze},
  editor       = {Matei Zaharia and
                  Gauri Joshi and
                  Yingyan (Celine) Lin},
  title        = {FlashInfer: Efficient and Customizable Attention Engine for {LLM}
                  Inference Serving},
  booktitle    = {Proceedings of the Eighth Conference on Machine Learning and Systems,
                  MLSys 2025, Santa Clara, CA, USA, May 12-15, 2025},
  publisher    = {OpenReview.net/mlsys.org},
  year         = {2025},
  url          = {https://openreview.net/forum?id=RXPofAsL8F},
  bibsource    = {dblp computer science bibliography, https://dblp.org}
}

@inproceedings{laskaridisMELTingPointMobile2024,
	author = {Laskaridis, Stefanos and Katevas, Kleomenis and Minto, Lorenzo and Haddadi, Hamed},
	title = {{MELTing} {Point}: {Mobile} {Evaluation} of {Language} {Transformers}},
	booktitle = {Proceedings of ACM MobiCom},
	year = {2024}
}

@article{wenAutoDroidV2BoostingSLMbased2025,
	title = {{AutoDroid}-{V2}: {Boosting} {SLM}-based {GUI} {Agents} via {Code} {Generation}},
	shorttitle = {{AutoDroid}-{V2}},
	urldate = {2025-09-03},
	journal = {arXiv preprint arXiv:2412.18116},
	author = {Wen, Hao and Tian, Shizuo and Pavlov, Borislav and Du, Wenjie and Li, Yixuan and Chang, Ge and Zhao, Shanhui and Liu, Jiacheng and Liu, Yunxin and Zhang, Ya-Qin and Li, Yuanchun},
	month = may,
	year = {2025},
}

@article{huangMNNAECSEnergyOptimization2025,
	title = {{MNN}-{AECS}: {Energy} {Optimization} for {LLM} {Decoding} on {Mobile} {Devices} via {Adaptive} {Core} {Selection}},
	shorttitle = {{MNN}-{AECS}},
	urldate = {2025-09-16},
	journal = {arXiv preprint arXiv:2506.19884},
	author = {Huang, Zhengxiang and Niu, Chaoyue and Wang, Zhaode and Xue, Jiarui and Zhang, Hanming and Wang, Yugang and Xin, Zewei and Jiang, Xiaotang and Lv, Chengfei and Wu, Fan and Chen, Guihai},
	month = jun,
	year = {2025},
}

@article{murthyMobileAIBenchBenchmarkingLLMs2024,
	title = {{MobileAIBench}: {Benchmarking} {LLMs} and {LMMs} for {On}-{Device} {Use} {Cases}},
	shorttitle = {{MobileAIBench}},
	urldate = {2025-09-24},
	journal = {arXiv preprint arXiv:2406.10290},
	author = {Murthy, Rithesh and Yang, Liangwei and Tan, Juntao and Awalgaonkar, Tulika Manoj and Zhou, Yilun and Heinecke, Shelby and Desai, Sachin and Wu, Jason and Xu, Ran and Tan, Sarah and Zhang, Jianguo and Liu, Zhiwei and Kokane, Shirley and Liu, Zuxin and Zhu, Ming and Wang, Huan and Xiong, Caiming and Savarese, Silvio},
	month = jun,
	year = {2024},
}

@article{xuFastOndeviceLLM2024,
	title = {Fast {On}-device {LLM} {Inference} with {NPUs}},
	urldate = {2025-10-15},
	journal = {arXiv preprint arXiv:2407.05858},
	author = {Xu, Daliang and Zhang, Hao and Yang, Liming and Liu, Ruiqi and Huang, Gang and Xu, Mengwei and Liu, Xuanzhe},
	month = dec,
	year = {2024},
}

@article{haoScalingLLMTestTime2025,
	title = {Scaling {LLM} {Test}-{Time} {Compute} with {Mobile} {NPU} on {Smartphones}},
	urldate = {2025-11-11},
	journal = {arXiv preprint arXiv:2509.23324},
	author = {Hao, Zixu and Wei, Jianyu and Wang, Tuowei and Huang, Minxing and Jiang, Huiqiang and Jiang, Shiqi and Cao, Ting and Ren, Ju},
	month = sep,
	year = {2025},
}

@inproceedings{luDemystifyingSmallLanguage2025,
	author = {Lu, Zhenyan and Li, Xiang and Cai, Dongqi and Yi, Rongjie and Liu, Fangming and Liu, Wei and Luan, Jian and Zhang, Xiwen and Lane, Nicholas D. and Xu, Mengwei},
	title = {Demystifying {Small} {Language} {Models} for {Edge} {Deployment}},
	booktitle = {Proceedings of ACL},
	year = {2025}
}

@inproceedings{wangLmMeterUnveilingRuntime2025,
	author = {Wang, Haoxin and Tu, Xiaolong and Ke, Hongyu and Chai, Huirong and Chen, Dawei and Han, Kyungtae},
	title = {lm-{Meter}: {Unveiling} {Runtime} {Inference} {Latency} for {On}-{Device} {Language} {Models}},
	booktitle = {Proceedings of ACM/IEEE SEC},
	year = {2025}
}

@article{zhangDissectingImpactMobile2025a,
	title = {Dissecting the {Impact} of {Mobile} {DVFS} {Governors} on {LLM} {Inference} {Performance} and {Energy} {Efficiency}},
	urldate = {2026-01-23},
	journal = {arXiv preprint arXiv:2507.02135},
	author = {Zhang, Zongpu and Dash, Pranab and Hu, Y. Charlie and Xu, Qiang and Li, Jian and Guan, Haibing},
	month = jul,
	year = {2025},
}

@inproceedings{chen_characterizing_2025,
	author = {Chen, Le and Feng, Dahu and Feng, Erhu and Wang, Yingrui and Zhao, Rong and Xia, Yubin and Xu, Pinjie and Chen, Haibo},
	title = {Characterizing {Mobile} {SoC} for {Accelerating} {Heterogeneous} {LLM} {Inference}},
	booktitle = {Proceedings of the ACM SOSP},
	year = {2025}
}

@inproceedings{wen_autodroid_2024,
	author = {Wen, Hao and Li, Yuanchun and Liu, Guohong and Zhao, Shanhui and Yu, Tao and Li, Toby Jia-Jun and Jiang, Shiqi and Liu, Yunhao and Zhang, Yaqin and Liu, Yunxin},
	title = {{AutoDroid}: {LLM}-powered {Task} {Automation} in {Android}},
	booktitle = {Proceedings of ACM MobiCom},
	year = {2024}
}

@inproceedings{guo_large_2025,
	author = {Guo, Qingzhe and Ouyang, Tu and Wang, An},
	title = {Large {Language} {Models} on {Mobile} {Devices}: {A} {Measurement} {Study} of {Single}- and {Multi}-{Instance} {Execution}},
	booktitle = {Proceedings of the 2nd {International} {Workshop} on {Edge} and {Mobile} {Foundation} {Models}},
	year = {2025}
}

@inproceedings{li_large_2024,
	author = {Li, Xiang and Lu, Zhenyan and Cai, Dongqi and Ma, Xiao and Xu, Mengwei},
	title = {Large {Language} {Models} on {Mobile} {Devices}: {Measurements}, {Analysis}, and {Insights}},
	booktitle = {Proceedings of the {Workshop} on {Edge} and {Mobile} {Foundation} {Models}},
	year = {2024}
}

@article{gptq,
	title = {{GPTQ}: Accurate Post-Training Quantization for Generative Pre-trained Transformers},
	journal = {arXiv preprint arXiv:2210.17323},
	author = {Frantar, Elias and Ashkboos, Saleh and Hoefler, Torsten and Alistarh, Dan},
	month = oct,
	year = {2022},
}

@article{awq,
	title = {{AWQ}: Activation-aware Weight Quantization for LLM Compression and Acceleration},
	journal = {arXiv preprint arXiv:2306.00978},
	author = {Lin, Ji and Tang, Jiaming and Tang, Haotian and Yang, Shang and Chen, Wei-Ming and Wang, Wei-Chen and Xiao, Guangxuan and Dang, Xingyu and Gan, Chuang and Han, Song},
	month = jun,
	year = {2023},
}

@article{sparsegpt,
	title = {{SparseGPT}: Massive Language Models Can Be Accurately Pruned in One-Shot},
	journal = {arXiv preprint arXiv:2301.00774},
	author = {Frantar, Elias and Alistarh, Dan},
	month = jan,
	year = {2023},
}

@article{flashattention,
	title = {{FlashAttention}: Fast and Memory-Efficient Exact Attention with IO-Awareness},
	journal = {arXiv preprint arXiv:2205.14135},
	author = {Dao, Tri and Fu, Daniel Y. and Ermon, Stefano and Rudra, Atri and Ré, Christopher},
	month = may,
	year = {2022},
}

@article{flashattention2,
	title = {{FlashAttention}-2: Faster Attention with Better Parallelism and Work Partitioning},
	journal = {arXiv preprint arXiv:2307.08691},
	author = {Dao, Tri},
	month = jul,
	year = {2023},
}

@article{vllmPagedAttention,
	title = {Efficient Memory Management for Large Language Model Serving with PagedAttention},
	journal = {arXiv preprint arXiv:2309.06180},
	author = {Kwon, Woosuk and Li, Zhuohan and Zhuang, Siyuan and Sheng, Ying and Zheng, Lianmin and Yu, Cody Hao and Gonzalez, Joseph E. and Zhang, Hao and Stoica, Ion},
	month = sep,
	year = {2023},
}

@article{h2o,
  author       = {Zhenyu Zhang and
                  Ying Sheng and
                  Tianyi Zhou and
                  Tianlong Chen and
                  Lianmin Zheng and
                  Ruisi Cai and
                  Zhao Song and
                  Yuandong Tian and
                  Christopher R{\'{e}} and
                  Clark W. Barrett and
                  Zhangyang Wang and
                  Beidi Chen},
  title        = {H\({}_{\mbox{2}}\)O: Heavy-Hitter Oracle for Efficient Generative
                  Inference of Large Language Models},
  journal      = {CoRR},
  volume       = {abs/2306.14048},
  year         = {2023},
  url          = {https://doi.org/10.48550/arXiv.2306.14048},
  doi          = {10.48550/ARXIV.2306.14048},
  eprinttype   = {arXiv},
  eprint       = {2306.14048},
  bibsource    = {dblp computer science bibliography, https://dblp.org}
}

@article{speculativeDecoding,
	title = {Fast Inference from Transformers via Speculative Decoding},
	journal = {arXiv preprint arXiv:2211.17192},
	author = {Leviathan, Yaniv and Kalman, Matan and Matias, Yossi},
	month = nov,
	year = {2022},
}

@inproceedings{mobile_foundation_as_firmware,
	author = {Yuan, Jinliang and Yang, Chen and Cai, Dongqi and Wang, Shihe and Yuan, Xin and Zhang, Zeling and Li, Xiang and Zhang, Dingge and Mei, Hanzi and Jia, Xianqing and Wang, Shangguang and Xu, Mengwei},
	title = {Mobile Foundation Model as Firmware},
	booktitle = {Proceedings of ACM MobiCom},
	year = {2024}
}

@article{lpspec2025,
	title = {{LP-Spec}: Leveraging {LPDDR} {PIM} for Efficient {LLM} Mobile Speculative Inference with Architecture-Dataflow Co-Optimization},
	journal = {arXiv preprint arXiv:2508.07227},
	author = {He, Siyuan and Zhu, Zhantong and He, Yandong and Jia, Tianyu},
	month = aug,
	year = {2025},
}

@inproceedings{song2024powerinfer,
	author = {Song, Yixin and Mi, Zeyu and Xie, Haotong and Chen, Haibo},
	title = {{PowerInfer}: Fast Large Language Model Serving with a Consumer-Grade {GPU}},
	booktitle = {Proceedings of the ACM SOSP},
	year = {2024}
}

@article{xue2024powerinfer,
	title = {{PowerInfer}-2: Fast Large Language Model Inference on a Smartphone},
	author = {Xue, Zhenliang and Song, Yixin and Mi, Zeyu and Zheng, Xinrui and Xia, Yubin and Chen, Haibo},
	journal = {arXiv preprint arXiv:2406.06282},
	year = {2024},
}

@inproceedings{xu2024wip,
  author       = {Daliang Xu and
                  Hao Zhang and
                  Liming Yang and
                  Ruiqi Liu and
                  Mengwei Xu and
                  Xuanzhe Liu},
  title        = {WiP: Efficient {LLM} Prefilling with Mobile {NPU}},
  booktitle    = {Proceedings of the Workshop on Edge and Mobile Foundation Models,
                  EdgeFM 2024, Minato-ku, Tokyo, Japan, June 3-7, 2024},
  pages        = {33--35},
  publisher    = {{ACM}},
  year         = {2024},
  url          = {https://doi.org/10.1145/3662006.3662066},
  doi          = {10.1145/3662006.3662066},
  bibsource    = {dblp computer science bibliography, https://dblp.org}
}

@article{yin2025dynamic,
	title = {Dynamic Sparse Attention on Mobile {SoCs}},
	author = {Yin, Wangsong and Xu, Daliang and Xu, Mengwei and Huang, Gang and Liu, Xuanzhe},
	journal = {arXiv preprint arXiv:2508.16703},
	year = {2025},
}

@ARTICLE{EdgeMoE,
author={Yi, Rongjie and Guo, Liwei and Wei, Shiyun and Zhou, Ao and Wang, Shangguang and Xu, Mengwei},
journal={ IEEE Transactions on Mobile Computing },
title={{ EdgeMoE: Empowering Sparse Large Language Models on Mobile Devices }},
year={2025},
volume={24},
number={08},
pages={7059-7073},
publisher={IEEE Computer Society},
address={Los Alamitos, CA, USA},
month=aug}

@ARTICLE{m2LLM,
  author={Liu, Kaiyuan and Zhou, Xiaobo and Li, Li},
  journal={IEEE Transactions on Parallel and Distributed Systems}, 
  title={m$^{2}$2LLM: A Multi-Dimensional Optimization Framework for LLM Inference on Mobile Devices}, 
  year={2025},
  volume={36},
  number={10},
  pages={2014-2029},
  doi={10.1109/TPDS.2025.3587445}}

@inproceedings{liu_visual_2023,
	author = {Liu, Haotian and Li, Chunyuan and Wu, Qingyang and Lee, Yong Jae},
	title = {Visual {Instruction} {Tuning}},
	booktitle = {Proceedings of NeurIPS},
	year = {2023}
}

@article{chung_scaling_2024,
  author       = {Hyung Won Chung and
                  Le Hou and
                  Shayne Longpre and
                  Barret Zoph and
                  Yi Tay and
                  William Fedus and
                  Yunxuan Li and
                  Xuezhi Wang and
                  Mostafa Dehghani and
                  Siddhartha Brahma and
                  Albert Webson and
                  Shixiang Shane Gu and
                  Zhuyun Dai and
                  Mirac Suzgun and
                  Xinyun Chen and
                  Aakanksha Chowdhery and
                  Alex Castro{-}Ros and
                  Marie Pellat and
                  Kevin Robinson and
                  Dasha Valter and
                  Sharan Narang and
                  Gaurav Mishra and
                  Adams Yu and
                  Vincent Y. Zhao and
                  Yanping Huang and
                  Andrew M. Dai and
                  Hongkun Yu and
                  Slav Petrov and
                  Ed H. Chi and
                  Jeff Dean and
                  Jacob Devlin and
                  Adam Roberts and
                  Denny Zhou and
                  Quoc V. Le and
                  Jason Wei},
  title        = {Scaling Instruction-Finetuned Language Models},
  journal      = {J. Mach. Learn. Res.},
  volume       = {25},
  pages        = {70:1--70:53},
  year         = {2024},
  url          = {https://jmlr.org/papers/v25/23-0870.html},
  bibsource    = {dblp computer science bibliography, https://dblp.org}
}

@article{xu_-device_2024_review,
	title = {On-{Device} {Language} {Models}: {A} {Comprehensive} {Review}},
	shorttitle = {On-{Device} {Language} {Models}},
	urldate = {2026-03-12},
	journal = {arXiv preprint arXiv:2409.00088},
	author = {Xu, Jiajun and Li, Zhiyuan and Chen, Wei and Wang, Qun and Gao, Xin and Cai, Qi and Ling, Ziyuan},
	month = sep,
	year = {2024},
}

@inproceedings{attention_is_all_you_need,
  author       = {Ashish Vaswani and
                  Noam Shazeer and
                  Niki Parmar and
                  Jakob Uszkoreit and
                  Llion Jones and
                  Aidan N. Gomez and
                  Lukasz Kaiser and
                  Illia Polosukhin},
  editor       = {Isabelle Guyon and
                  Ulrike von Luxburg and
                  Samy Bengio and
                  Hanna M. Wallach and
                  Rob Fergus and
                  S. V. N. Vishwanathan and
                  Roman Garnett},
  title        = {Attention is All you Need},
  booktitle    = {Advances in Neural Information Processing Systems 30: Annual Conference
                  on Neural Information Processing Systems 2017, December 4-9, 2017,
                  Long Beach, CA, {USA}},
  pages        = {5998--6008},
  year         = {2017},
  url          = {https://proceedings.neurips.cc/paper/2017/hash/3f5ee243547dee91fbd053c1c4a845aa-Abstract.html},
  bibsource    = {dblp computer science bibliography, https://dblp.org}
}

@inproceedings{ainslie-etal-2023-gqa_kvcache,
	author = "Ainslie, Joshua  and
      Lee-Thorp, James  and
      de Jong, Michiel  and
      Zemlyanskiy, Yury  and
      Lebron, Federico  and
      Sanghai, Sumit",
	title = "{GQA}: Training Generalized Multi-Query Transformer Models from Multi-Head Checkpoints",
	booktitle = {Proceedings of EMNLP},
	year = "2023"
}

@misc{noauthor_qualcomm_nodate_qnn,
	title = {Qualcomm {AI} {Engine} {Direct} {SDK} {\textbar} {Qualcomm} {Developer}},
	url = {https://www.qualcomm.com/developer/software/qualcomm-ai-engine-direct-sdk},
	author       = {{Qualcomm Technologies, Inc.}},
	language = {en},
	urldate = {2026-03-12},
}

@misc{noauthor_hexagon_nodate,
	title = {Hexagon {NPU} {SDK} {\textbar} {Qualcomm} {Developer}},
	author       = {{Qualcomm Technologies, Inc.}},
	url = {https://www.qualcomm.com/developer/software/hexagon-npu-sdk},
	language = {en},
	urldate = {2026-03-12},
}

@misc{oneplus_qpt_kernel,
  author       = {{OnePlusOSS}},
  title        = {{android\_kernel\_oneplus\_sm8850}: Qualcomm power telemetry and powercap driver sources},
  year         = {2026},
  howpublished = {\url{https://github.com/OnePlusOSS/android_kernel_oneplus_sm8850/tree/6504e3d0385a951de4848bc81ce19ce8f8145dbe/drivers/powercap/qcom}},
  note         = {Commit 6504e3d0385a951de4848bc81ce19ce8f8145dbe, accessed 2026-03-13}
}

@article{xuEdgeLLMFastOnDevice2025,
	title = {{EdgeLLM}: {Fast} {On}-{Device} {LLM} {Inference} {With} {Speculative} {Decoding}},
	volume = {24},
	shorttitle = {{EdgeLLM}},
	language = {en-US},
	number = {4},
	urldate = {2025-05-22},
	journal = {IEEE Transactions on Mobile Computing},
	author = {Xu, Daliang and Yin, Wangsong and Zhang, Hao and Jin, Xin and Zhang, Ying and Wei, Shiyun and Xu, Mengwei and Liu, Xuanzhe},
	month = apr,
	year = {2025},
	pages = {3256--3273},
}

@software{llamacpp,
  author = {Georgi Gerganov},
  title  = {llama.cpp},
  year   = {2023},
  url    = {https://github.com/ggml-org/llama.cpp}
}

@software{mlc_llm,
  author = {{MLC Team}},
  title  = {MLC-LLM},
  year   = {2023},
  url    = {https://github.com/mlc-ai/mlc-llm}
}

@article{jiang2020mnn,
  author       = {Xiaotang Jiang and Huan Wang and Yiliu Chen and Ziqi Wu and
                  Lichuan Wang and Bin Zou and Yafeng Yang and Zongyang Cui and
                  Yu Cai and Tianhang Yu and Chengfei Lv and Zhihua Wu},
  title        = {MNN: A Universal and Efficient Inference Engine},
  journal      = {arXiv preprint arXiv:2002.12418},
  year         = {2020}
}

@misc{mllm,
  title = {mllm: fast and lightweight multimodal LLM inference engine for mobile and edge devices},
  author = {Rongjie Yi and Xiang Li and Zhenyan Lu and Hao Zhang and Daliang Xu and Liming Yang and Weikai Xie and Chenghua Wang and Yun Ma and Xuanzhe Liu and Mengwei Xu},
  year = {2023},
  publisher = {mllm Team},
  url = {https://github.com/UbiquitousLearning/mllm}
}

@misc{qualcomm_genie_extensions,
  author       = {{Qualcomm Technologies, Inc.}},
  title        = {Gen AI Inference Extensions},
  year         = {2026},
  howpublished = {\url{https://www.qualcomm.com/developer/software/gen-ai-inference-extensions}},
  note         = {Accessed: 2026-03-14}
}

@inproceedings{10.1109/ISCA59077.2024.00019,
  author       = {Pratyush Patel and
                  Esha Choukse and
                  Chaojie Zhang and
                  Aashaka Shah and
                  {\'{I}}{\~{n}}igo Goiri and
                  Saeed Maleki and
                  Ricardo Bianchini},
  title        = {Splitwise: Efficient Generative {LLM} Inference Using Phase Splitting},
  booktitle    = {51st {ACM/IEEE} Annual International Symposium on Computer Architecture,
                  {ISCA} 2024, Buenos Aires, Argentina, June 29 - July 3, 2024},
  pages        = {118--132},
  publisher    = {{IEEE}},
  year         = {2024},
  url          = {https://doi.org/10.1109/ISCA59077.2024.00019},
  doi          = {10.1109/ISCA59077.2024.00019},
  bibsource    = {dblp computer science bibliography, https://dblp.org}
}

@inproceedings{amini-etal-2019-mathqa,
    title = "{M}ath{QA}: Towards Interpretable Math Word Problem Solving with Operation-Based Formalisms",
    author = "Amini, Aida  and
      Gabriel, Saadia  and
      Lin, Shanchuan  and
      Koncel-Kedziorski, Rik  and
      Choi, Yejin  and
      Hajishirzi, Hannaneh",
    booktitle = "Proceedings of the 2019 Conference of the North {A}merican Chapter of the Association for Computational Linguistics: Human Language Technologies, Volume 1 (Long and Short Papers)",
    month = jun,
    year = "2019",
    publisher = "Association for Computational Linguistics",
}

@inproceedings{wang2024rolellm,
  author       = {Noah Wang and
                  Zhongyuan Peng and
                  Haoran Que and
                  Jiaheng Liu and
                  Wangchunshu Zhou and
                  Yuhan Wu and
                  Hongcheng Guo and
                  Ruitong Gan and
                  Zehao Ni and
                  Jian Yang and
                  Man Zhang and
                  Zhaoxiang Zhang and
                  Wanli Ouyang and
                  Ke Xu and
                  Wenhao Huang and
                  Jie Fu and
                  Junran Peng},
  editor       = {Lun{-}Wei Ku and
                  Andre Martins and
                  Vivek Srikumar},
  title        = {RoleLLM: Benchmarking, Eliciting, and Enhancing Role-Playing Abilities
                  of Large Language Models},
  booktitle    = {Findings of the Association for Computational Linguistics, {ACL} 2024,
                  Bangkok, Thailand and virtual meeting, August 11-16, 2024},
  series       = {Findings of {ACL}},
  volume       = {{ACL} 2024},
  pages        = {14743--14777},
  publisher    = {Association for Computational Linguistics},
  year         = {2024},
  url          = {https://doi.org/10.18653/v1/2024.findings-acl.878},
  doi          = {10.18653/V1/2024.FINDINGS-ACL.878},
  bibsource    = {dblp computer science bibliography, https://dblp.org}
}

@article{bai2024longbench2,
  title = {{LongBench v2}: Towards Deeper Understanding and Reasoning on Realistic Long-context Multitasks},
  author = {Bai, Yushi and Tu, Shangqing and Zhang, Jiajie and Peng, Hao and Wang, Xiaozhi and Lv, Xin and Cao, Shulin and Xu, Jiazheng and Hou, Lei and Dong, Yuxiao and Tang, Jie and Li, Juanzi},
  journal = {arXiv preprint arXiv:2412.15204},
  year = {2024},
  url = {https://arxiv.org/abs/2412.15204}
}

@article{benazir2025benchmarkingapplesilicon,
  title={Benchmarking and Characterization of Large Language Model Inference on Apple Silicon},
  author={Benazir, Afsara and Lin, Felix Xiaozhu},
  journal={Proceedings of the ACM on Measurement and Analysis of Computing Systems},
  volume={9},
  number={3},
  pages={1--26},
  year={2025},
  publisher={ACM New York, NY, USA}
}

@article{bengio2003neuralprobabilistic,
  author = {Bengio, Yoshua and Ducharme, R\'{e}jean and Vincent, Pascal and Jauvin, Christian},
  title = {A Neural Probabilistic Language Model},
  journal = {Journal of Machine Learning Research},
  volume = {3},
  pages = {1137--1155},
  year = {2003},
  url = {https://jmlr.org/papers/v3/bengio03a.html}
}

@inproceedings{merity2017pointersentinel,
  author = {Merity, Stephen and Xiong, Caiming and Bradbury, James and Socher, Richard},
  title = {Pointer Sentinel Mixture Models},
  booktitle = {International Conference on Learning Representations},
  year = {2017},
  url = {https://openreview.net/forum?id=Byj72udxe}
}

@article{executorch2026,
  title = {{ExecuTorch} - A Unified {PyTorch} Solution to Run {AI} Models On-Device},
  author = {Nachin, Mergen and Desai, Digant and Jia, Sicheng Stephen and Lai, Chen and Liu, Mengwei and Szwejbka, Jacob and Alvarez, Raziel and Ascani, RJ and Bort, Dave and Candales, Manuel and others},
  journal = {arXiv preprint arXiv:2605.08195},
  year = {2026},
  url = {https://arxiv.org/abs/2605.08195}
}

@inproceedings{xin2026sereno,
  author = {Tong Xin and Xinrui Shi and Mingkai Dong and Zeyu Mi},
  title = {Inference in the Shadows: Taming Memory Bandwidth Contention in Mobile {LLM} Inference with Sereno},
  booktitle = {20th USENIX Symposium on Operating Systems Design and Implementation (OSDI 26)},
  year = {2026},
  isbn = {978-1-939133-55-7},
  address = {Seattle, WA},
  pages = {2299--2318},
  url = {https://www.usenix.org/conference/osdi26/presentation/xin},
  publisher = {USENIX Association},
  month = jul
}
\appendix

\end{document}